\documentclass[a4paper,12pt]{article}
\usepackage{amsmath,amsthm,amsfonts,amssymb,bm,mathrsfs,commath}
\usepackage[scale=0.75]{geometry}
\usepackage{xcolor,graphicx}
\usepackage{enumitem}
\setlist{nosep}

\usepackage{comment}
\usepackage{natbib}
\usepackage{placeins,sgame} 

\usepackage{setspace}
\usepackage{hyperref}

\hypersetup{colorlinks=true,
           citecolor=blue,
           linkcolor=blue,
           urlcolor=blue,
           bookmarksopen=true,
           pdfstartview={XYZ null null 1.30},
           pdfpagelayout={SinglePage}
}

\allowdisplaybreaks
\newtheorem{defn}{Definition}
\newtheorem{thm}{Theorem}
\newtheorem{lem}{Lemma}

\newtheorem{prop}{Proposition}
\newtheorem{rmk}{Remark}
\newtheorem{exam}{Example}
\newtheorem{coro}{Corollary}
\newtheorem{claim}{Claim}

\newcommand{\bE}{\mathbf{E}}

\newcommand{\bP}{\mathbf{P}}
\newcommand{\bR}{\mathbb{R}}

\newcommand{\cB}{\mathcal{B}}

\newcommand{\cD}{\mathcal{D}}

\newcommand{\cF}{\mathcal{F}}
\newcommand{\cG}{\mathcal{G}}

\newcommand{\cM}{\mathcal{M}}

\newcommand{\cR}{\mathcal{R}}

\newcommand{\cT}{\mathcal{T}}

\DeclareMathOperator{\rmd}{d\!}

\DeclareMathOperator{\supp}{supp}
\DeclareMathOperator*{\argmax}{arg\,max}

\begin{document}

\title{Characterization of equilibrium existence and purification in general Bayesian games\footnote{Some of the results reported here were presented at the 2017 Asian Meeting of the Econometric Society, The Chinese University of Hong Kong; and the 17th Society for the Advancement of Economic Theory Conference, Faro. We thank the participants for their helpful comments. Part of the results in this paper were included in Chapter 3 of Yishu Zeng's 2016 Ph.D dissertation (\citet{Zeng2016}). } }

\author{Wei~He\thanks{Department of Economics, The Chinese University of Hong Kong, Hong Kong. E-mail: hewei@cuhk.edu.hk}
\and
Xiang~Sun\thanks{Center for Economic Development Research, Economics and Management School, Wuhan University, 299 Bayi Road, Wuhan, 430072, China. Email: xiangsun.econ@gmail.com}
\and
Yeneng~Sun\thanks{Departments of Economics and Mathematics, National University of Singapore, Singapore, 119076. Email: ynsun@nus.edu.sg}
\and
Yishu~Zeng\thanks{Department of Economics, University of Michigan, Ann Arbor, United States. Email: zengyish@umich.edu}}

\date{This version: \today}

\maketitle

\begin{abstract}
This paper studies Bayesian games with general action spaces, correlated types and interdependent payoffs. We introduce the condition of ``decomposable coarser payoff-relevant information'', and show that this condition is both sufficient and necessary for the existence of pure-strategy equilibria and purification from behavioral strategies.
As a consequence of our purification method,
a new existence result on pure-strategy equilibria is also obtained for discontinuous Bayesian games.
Illustrative applications of our results to oligopolistic competitions and all-pay auctions are provided.

\end{abstract}

{\singlespace{\textbf{Keywords}: General Bayesian game; Decomposable coarser payoff-relevant information; Pure-strategy equilibrium; Existence; Purification}}

\clearpage
\clearpage


\section{Introduction}
\label{sec:intro}

Bayesian games (\`{a} la \cite{Harsanyi}), where players have incomplete information about certain aspects of the environment, arise naturally in numerous real-life situations. Many economic applications are most conveniently formulated as Bayesian games with infinitely many choices. Consider, for instance, several bidders competing for a single object in an auction, where each bidder only knows her private information and chooses bids from some bid space. While it is true that the bid spaces in practical situations are in fact discrete, studying the setting with continuum bid spaces has appealing advantages. On the one hand, solving the case with finite bid spaces may involve complicated combinatorial arguments that need multiple steps of approximation. On the other hand,  one often needs to use calculus to characterize equilibria and to conduct the comparative statics analysis in the setting of continuum bid spaces.\footnote{Note that previous studies on auctions---one of the most successful applications of Bayesian games, generally allow bidders to choose from continuum bid spaces; see, for example, \cite{Krishna2009}.} The consideration of general action spaces allows one not only to simplify the analysis of Bayesian games, but also to uncover interesting results that cannot be found in the discrete framework.

The fundamental question of equilibrium existence for Bayesian games with general action spaces has been extensively studied in the literature. To ensure that players' expected payoffs are continuous in strategy profiles, \cite{MW1985} worked with continuous payoffs and assumed the absolute continuity (AC) condition on the information structure. They provided an existence result for behavioral-strategy equilibria in general Bayesian games.\footnote{The AC condition is widely satisfied in economic applications, and plays an essential role in the proof of the equilibrium existence results in subsequent works. The recent studies on Bayesian games with discontinuous payoffs (\textit{e.g.}, \cite{CM2018} and \cite{PY2019}) present various equilibrium existence results by proposing different payoff security-type conditions under the same AC condition on the information structure. Without the AC condition, a behavioral-strategy equilibrium may not exist even in the finite-action setting: \cite{Simon2003} constructed a Bayesian game without any equilibrium;  \cite{Hellman2014} provided a simpler example without any approximate  equilibrium; and \cite{FM2017} presented an example associated with the universal type structure that does
not have any equilibrium.}

Despite its wide use in the studies on Bayesian games, the notion of behavioral strategy has been criticized for various reasons. As noted in \cite{RR1982} and \cite{MW1985}, one rarely observes that individuals make decisions by using randomization devices in practical situations. In addition, economic applications of games with incomplete information often focus on pure-strategy equilibrium. If one adopts pure-strategy equilibrium as the solution concept, then several basic questions could be asked. Do pure-strategy equilibria generally exist in Bayesian games? If pure-strategy equilibria do exist, can players obtain the same equilibrium payoffs (and thus the same social welfare) as those based on behavioral-strategy equilibria; that is, will focusing on pure-strategy equilibria be without loss of generality? The answers to these questions are negative.\footnote{In Bayesian games with finitely many actions, the existence of pure-strategy equilibria has been well established. Based on \cite{DWW1951a}'s purification result, it is straightforward to obtain the existence of pure-strategy equilibria in the setting with private values and (conditionally) independent types; see \cite{RR1982}, \cite{MW1985}, and \cite{KRS2006}. \cite{KZ2018} worked with finite-action Bayesian games with payoff-irrelevant correlated types. \cite{HS2019} characterized several important properties of pure-strategy equilibria via the ``coarser inter-player information'' condition in finite-action Bayesian games. \cite{HL2017} proved the equilibrium existence result for Bayesian games with purely atomic types under a smoothness condition. \cite{BK2019} studied games with incomplete information and non-expected utilities.} \cite{KRS1999} presented a simple two-player Bayesian game with private values and continuous payoffs,\footnote{A player is said to have private values (resp. interdependent payoff) if her payoff function depends on her own type (resp. all players' types) and the action profile. } where both players' action spaces are $[-1, 1]$, and the common prior is the uniform distribution on the square $[0,1] \times [0,1]$ (\textit{i.e.}, the two players have independent types with uniform distribution on their individual type space $[0,1]$). They pointed out that this Bayesian game does not possess any pure-strategy equilibrium, while behavioral-strategy equilibria always exist.\footnote{The AC condition trivially holds in this example due to independent types. In addition, the condition of ``coarser inter-player information'' in \cite{HS2019} is also satisfied given the private values and independent types. The nonexistence of pure-strategy equilibria in this example indicates that the results in \cite{HS2019}  cannot be applied to the infinite-action setting.} Then the question is, under what kind of suitable conditions, pure-strategy equilibria exist in Bayesian games with general action spaces.

We aim to address the above questions in this paper, and attempt to provide practitioners with a ``toolkit'' of relatively simple conditions that are useful in proving the existence of pure-strategy equilibria in applied works. The focus is on general Bayesian games, which allow players' actions to be infinite, types to be correlated, and payoffs to be interdependent. Towards this end, we introduce the condition of ``decomposable coarser payoff-relevant information'' (DCPI in short) in the sense
that the density-weighted payoff of each player is decomposed into a sum of finitely many components with each component being the product of a ``coarser'' action-relevant part and an action-irrelevant part.

Under DCPI, Theorem~\ref{thm:exist} establishes the existence of pure-strategy equilibria for Bayesian games with general action spaces.
\citet[p.~236]{FT1991} considered Bayesian games with continuous payoffs on compact action spaces, private values and conditionally independent types.
If one imposes the condition that each player's payoff-relevant private information is coarser than her full private information given any non-trivial event, then DCPI is satisfied. Thus, Theorem~\ref{thm:exist} implies the existence of pure-strategy equilibria in the particular setting; see Corollary \ref{coro-private} below.\footnote{The counterexample in \cite{KRS1999} as mentioned above indicates that such an existence result may fail without the coarseness condition.}
Remark 1 in \citet[p.~236]{FT1991} also pointed out
the need to find regularity conditions for working with pure-strategy equilibria in general Bayesian games without the restriction of conditionally independent types and private values.
The DCPI condition serves this purpose by allowing us to consider Bayesian games with general action spaces, interdependent payoffs and correlated types as in Theorem~\ref{thm:exist}. Furthermore, DCPI is shown to be a minimal regularity condition in the sense that it is satisfied if pure-strategy equilibria always exist in general Bayesian games; see Proposition \ref{prop:exist}. Example \ref{exam cond independence} provides an illustrative application of Theorem~\ref{thm:exist} to oligopolistic competitions.

In Theorem~\ref{thm:purification}, we present a purification principle (called conditional purification) relating behavioral strategies to pure strategies. A conditional purification of a behavioral-strategy profile preserves the same expected payoffs and action distributions conditional on any non-trivial event in players' types, and thus also preserves the equilibrium property. We show that
the DCPI condition is both sufficient and necessary for the existence of conditional purification in general Bayesian games. In this sense, DCPI appears to be an appropriate condition showing that focusing on pure strategies is without loss of generality. To demonstrate the usefulness of the purification result, we apply it to discontinuous Bayesian games, which arise naturally in economic applications. For instance, auctions and price competitions are typical examples where players' payoffs are not continuous when a tie occurs. Building on the result of \cite{Reny1999} for normal form games, \cite{CM2018} and \cite{HY2016} established the equilibrium existence in behavioral strategies in the incomplete information setting. We present a new equilibrium existence result in pure strategies for discontinuous Bayesian games in Proposition~\ref{prop:discontinuous}, and
demonstrate its applicability to all-pay auctions in Example \ref{exam:allpay}.

There is a sizable literature studying the existence of pure-strategy equilibria and the purification method in Bayesian games. A major approach is to impose rich structures on the type spaces. \cite{KS1999} showed that a pure-strategy equilibrium exists by modelling players' type spaces as atomless Loeb spaces.\footnote{Loeb spaces were first introduced in \cite{Loeb1975}; see \cite{LW2015} for the construction.} The equilibrium existence result was further generalized to the setting with saturated probability spaces as the type spaces; see \cite{WZ2012} and \cite{KZ2014}. \cite{LS2006} obtained purification results with general action spaces by considering atomless Loeb spaces. The purification method was further studied in \cite{Podczeck2009} based on saturated probability spaces. \cite{HS2014} relaxed the assumption of saturated probability spaces by the relative diffuseness condition for both the equilibrium existence and purification results. All these papers study Bayesian games with private values and (conditionally) independent types. In this paper, we allow for interdependent payoffs and correlated types, and dispense with the additional rich structures on the type spaces---Loeb spaces/saturated probability spaces/relative diffuseness.\footnote{Working with Loeb spaces/saturated probability spaces excludes Polish spaces as the type spaces, which are the widely used private type spaces in economic games. Our results do not have this restriction.} Therefore, our results cover the results in all those papers.

Another stream of literature studies the existence of pure-strategy equilibria in Bayesian games by assuming order structures on the payoffs. \cite{Vives1990} presented an existence result of pure-strategy equilibria in supermodular games. By assuming the Spence-Mirrlees single crossing property, \cite{Athey2001} proved that a monotone pure-strategy equilibrium exists, which was later extended to the setting with multidimensional and partially ordered type/action spaces in \cite{McAdams2003}. \cite{Reny2011} further generalized the results in these two papers by allowing the action spaces to be compact locally complete metric semilattices and the type spaces to be partially ordered probability spaces.

The paper is organized as follows. Section~\ref{sec:model} describes the formulation of general Bayesian games. Section~\ref{sec:cdpi} introduces the condition of DCPI. Section~\ref{sec:exist} shows that this condition is sufficient and necessary for the existence of pure-strategy equilibria. In Section~\ref{sec:purification}, we characterize the existence of conditional purifications via DCPI, and present a new equilibrium existence result for discontinuous Bayesian games. The proofs are collected in Section~\ref{sec:appendix}.


\section{General Bayesian games}
\label{sec:model}

We start by describing the model in terms of the players, types, actions, payoffs, strategies, and the equilibrium notion.

\paragraph{Players}
The set of players is $I = \{1, 2, \ldots, n\}$ with $n \ge 2$.

\paragraph{Types}
For each $i \in I$, player~$i$ observes a private type $t_i$, whose value lies in some measurable space $(T_i, \cT_i)$. The set $T = \prod_{i \in I} T_i$ collects all the type profiles, which is endowed with $\cT = \otimes_{i \in I} \cT_i$. Let $\lambda$ be the common prior in the game, which is a probability measure on the space of type profiles $(T, \cT)$.

For each $i \in I$, the marginal of the common prior $\lambda$ on player~$i$'s private type space $(T_i, \cT_i)$ is denoted by $\lambda_i$. We shall work with the condition that $\lambda$ admits a density with respect to $\otimes_{i \in I} \lambda_i$; that is, types can be correlated as long as the common prior $\lambda$ is absolutely continuous with respect to $\otimes_{i \in I} \lambda_i$ with the corresponding Radon-Nikodym derivative $q$.\footnote{Let $(T, \cT, \lambda)$ be a probability space. A finite measure $\nu$ is said to be absolutely continuous with respect to $\lambda$ if for any $D \in \cT$, $\lambda(D) = 0$ implies $\nu(D) = 0$. In this case, there exists a ($\lambda$-almost) unique $\lambda$-integrable function $q$ such that $\nu(D) = \int_D q(t) \lambda(\dif t)$ for any $D \in \cT$. Such a function $q$ is called the Radon-Nikodym derivative of $\nu$ with respect to $\lambda$; see Theorem~13.18 in \cite{AB2006}.} When the marginals are atomless, this assumption simply means that $t_i$ and $t_\ell$ are not too dependent for distinct players $i$ and $\ell$. For example, it excludes the case of perfect correlation for $t_i$ and $t_\ell$. However, if $t_i$ is multi-dimensional for some player~$i$, this assumption puts no restriction on how different components of $t_i$ are related.\footnote{That is, if $t_i = (t_{i1}, \ldots, t_{ik})$ is multi-dimensional, then different components $t_{i1}$, ..., $t_{ik}$ of $t_i$ can be arbitrarily correlated with each other. } As usual, the notation $-i$ denotes all the players except player~$i$, and $\lambda_{-i} = \otimes_{\ell \ne i} \lambda_\ell$.

\paragraph{Actions}
After observing the private type, player~$i \in I$ chooses an action from the action space $A_i$. The set $A_i$ is a nonempty and compact metric space endowed with the Borel $\sigma$-algebra $\cB(A_i)$. We denote the set of all action profiles by $A = \prod_{i \in I} A_i$.

\paragraph{Payoffs}
For each player~$i \in I$, the payoff $u_i$ depends on the action profile $a \in A$ as well as the type profile $t \in T$. We shall assume that each $u_i$ is a well-behaved mapping. Specifically,
\begin{itemize}
\item $u_i$ is jointly measurable on $T \times A$;
\item $u_i$ is integrably bounded in the sense that there is a real-valued integrable mapping $h_i$ on $(T, \cT, \lambda)$ with $|u_i(a,t)| \le h_i(t)$ for all $(a,t) \in A \times T$.
\end{itemize}
Note that players' payoffs are allowed to be interdependent.

\paragraph{Strategies}
For player~$i\in I$, let $\cM(A_i)$ be the space of Borel probability measures on $A_i$ endowed with the topology of weak convergence. For each player~$i \in I$, a behavioral strategy is a measurable mapping from the private type space $(T_i, \cT_i)$ to $\cM(A_i)$.\footnote{A distributional strategy of player~$i$ is a probability measure on the product of her type and action spaces, with the marginal being $\lambda_i$ on $(T_i, \cT_i)$. It is clear that every behavioral strategy corresponds to a natural distributional strategy, and every distributional strategy induces an equivalent class of behavioral strategies. We focus on behavioral strategies for simplicity, but all the results can be easily extended to distributional strategies.} Let $L_i^{\cT_i}$ be the set of all behavioral strategies for player~$i$. Denote $L^{\cT} = \prod_{i \in I} L_i^{\cT_i}$. Similarly, we can define a pure strategy as a measurable mapping from $(T_i, \cT_i)$ to $A_i$, which can be viewed as a behavioral strategy by taking it as a Dirac measure for all $t_i \in T_i$.

\paragraph{Equilibria}
Given a strategy profile $h = (h_1, h_2, \ldots, h_n)$, player~$i$'s expected payoff is
\begin{align*}
U_i(h)  & = \int_T \int_A  u_i(a,t)  \cdot  {\textstyle \prod\limits_{\ell \in I}} h_\ell(t_\ell; \dif a_\ell)  \lambda(\dif t)  \\
        & = \int_T \int_A  u_i(a,t)  \cdot  q(t)  \cdot  {\textstyle \prod\limits_{\ell \in I}} h_\ell(t_\ell; \dif a_\ell)  \mathop{\otimes}_{\ell \in I} \lambda_\ell(\dif t_\ell).
\end{align*}
A \emph{behavioral-strategy equilibrium} (resp. \emph{pure-strategy equilibrium}) is a behavioral-strategy profile (resp. pure-strategy profile) $h^*=(h^*_1, h_2^*, \ldots, h^*_n)$ such that $h^*_i$ maximizes $U_i(h_i, h^*_{-i})$ for each player $i \in I$.


\section{Decomposable coarser payoff-relevant information}
\label{sec:cdpi}

In this section, we propose a condition to describe the difference between the information conveyed via the types and the information conveyed via the payoff-relevant components of the types.\footnote{Players' types may contain not only the payoff-relevant information, but also the payoff-irrelevant information. For example, the beliefs may differ given distinct types, and thus one could be able to design mechanisms eliciting players' beliefs about others; see \cite{CM1988}, \cite{HN2006}, \cite{CX2013}, and \cite{Guo2019} for more discussions.}

In the definition of the expected payoff $U_i$ for player~$i$, it is strategically equivalent for each player~$i$ to view $u_i(a,t) \cdot q(t)$ as the payoff and $\otimes_{\ell \in I} \lambda_\ell$ as the prior. Hereafter, we shall work with the \emph{density-weighted payoff}
$$ w_i(a, t) = u_i(a, t) \cdot q(t) $$
for each player~$i \in I$, action profile~$a \in A$, and type profile~$t\in T$. For a strategy profile $f = (f_1, f_2, \ldots, f_n)$, player~$i$'s expected payoff can be rewritten as
$$ U_i(f) = \int_T \int_A  w_i(a, t) \cdot {\textstyle\prod\limits_{\ell \in I}} f_\ell(t_\ell; \dif a_\ell)  \mathop{\otimes}\limits_{\ell \in I} \lambda_\ell (\dif t_\ell). $$

Before stating the key condition of ``decomposable coarser payoff-relevant information,'' we first need to define the notion of ``nowhere equivalence.'' Let $(\Omega, \hat\cT, \bP)$ be an atomless finite positive measure space, and $\hat\cF$ a sub-$\sigma$-algebra of $\hat\cT$. For a set $D \in \hat\cT$ with $\bP(D) > 0$, let $\hat\cF^D$ (resp. $\hat\cT^D$) be the restricted $\sigma$-algebra $\{ D \cap D' \mid D' \in \hat\cF\}$ (resp. $\{ D \cap D' \mid D' \in \hat\cT \}$) on $D$. The $\sigma$-algebra $\hat\cT$ is said to be nowhere equivalent to $\hat\cF$ under $\bP$ if the strong completion of $\hat\cF^D$ in $\hat\cT^D$ under $\bP$ is not equal to $\hat\cT^D$ for any $D \in \hat\cT$ of positive measure.\footnote{The strong completion of $\hat\cF^D$ in $\hat\cT^D$ under $\bP$ is the collection of all sets in the form $E \triangle E_0$, where $E \in \hat\cF^D$ and $E_0$ is a $\bP$-null set in $\hat\cT^D$, and $E \triangle E_0$ denotes the symmetric difference $(E \setminus E_0) \cup (E_0 \setminus E)$.}

Let $\cF_i$ be a sub-$\sigma$-algebra of $\cT_i$. Throughout the paper, we assume that $(T_i, \cF_i, \lambda_i)$ is atomless for each $i \in I$.\footnote{In economic applications of Bayesian games, the type spaces are often modelled by intervals or rectangles with density functions, which are atomless probability spaces. Furthermore, it is natural to consider infinite types in Bayesian games, as the set of belief types may have the cardinality of the continuum; see, for example, the discussions in \cite{BD1993} and \cite{Hammond2004}.} We introduce the definition of decomposable coarser payoff-relevant information as follows.
\begin{defn}
\label{defn:cdpi}
\rm
A Bayesian game is said to have \emph{decomposable coarser payoff-relevant information} (DCPI hereafter) if each $\cT_i$ is nowhere equivalent to $\cF_i$ under $\lambda_i$, and there exists a positive integer $J$ such that for player $i \in I$,
$$w_i(a, t) = \sum_{j=1}^J \biggl[ w^j_i(a, t) \cdot \textstyle{\prod\limits_{\ell \in I}} \rho^j_\ell(t_\ell) \biggr], $$
where for $j = 1, 2, \ldots, J$, (1) $w^j_i(a, \cdot)$ is $\otimes_{\ell \in I} \cF_\ell$-measurable and $\otimes_{\ell \in I} \lambda_\ell$-integrable for each $a \in A$, (2) $\rho^j_\ell$ is nonnegative and integrable on $(T_\ell, \cT_\ell, \lambda_\ell)$ for each $\ell \in I$.
\end{defn}

In a Bayesian game with DCPI, players have interdependent payoffs and correlated types, as players' payoff functions can depend on the types of each other, and the density function can be non-trivial. DCPI means that the density-weighted payoff of each player is decomposed into a sum of finitely many components with each component being the product of a ``coarser'' action-relevant part and an action-irrelevant part.
When $J = 1$ and $\rho^1_1 \equiv 1$, $w^J_i$ is simply the density-weighted payoff $w_i$. If the density-weighted payoffs are required to satisfy the DCPI condition for $J = 1$, then  the functions $\{w_i\}_{i \in I}$ need to be measurable with respect to $\otimes_{\ell \in I} \cF_\ell$. In various economic applications of Bayesian games, players are often assumed to have conditionally independent types.\footnote{For example, auctions with conditionally independent types have been studied in both the empirical and theoretical literatures; see \cite{LPV2000} and \cite{Athey2001}.} As demonstrated by Example~\ref{exam cond independence} below, DCPI would fail for a simple two-player Bayesian game with conditionally independent payoff-irrelevant types if $J$ is required to be $1$. To address this issue, we allow $J$ to be any positive integer. Indeed, Corollary \ref{coro-interdependence} below presents a general model of Bayesian games with interdependent payoffs and conditionally independent types that satisfy the DCPI condition.
In Sections~\ref{subsec:application} and \ref{subsec:discontinuous} below, we also consider specific Bayesian games with conditionally independent types in the settings of oligopolistic competitions and all-pay auctions, and show that these games satisfy DCPI for a general $J \ge 1$.

Below, we present an example of a two-player Bayesian game to illustrate the DCPI condition.\footnote{This example is a variation of Example~1 in \cite{HS2019}.}

\begin{exam}\label{exam cond independence}
There are two players, $I = \{1, 2\}$. For $i \in I$, $T_i = [0,1]$ endowed with the Borel $\sigma$-algebra $\cT_i = \cB([0,1])$. The action space of each player is $[0,1]$. Player~$i$'s payoff function $u_i$ is bounded, continuous, and only depends on the action profile. The set of common states $T_0 = \{t_{01}, t_{02}\}$. The two common states are drawn with equal probability, which are unobservable to both players. At $t_{01}$, a pair $(t_1, t_2) \in T_1 \times T_2$ is drawn following the uniform distribution $\tilde{\lambda} = \eta \otimes \eta$. At $t_{02}$, $(t_1, t_2)$ is drawn under the distribution $\hat{\lambda}$, which has the density $6t_1 t^2_2$ with respect to the uniform distribution. The space of type profiles is $T = T_1 \times T_2$, and the common prior is $\lambda = \frac{1}{2}\left(\tilde{\lambda} + \hat{\lambda}\right)$.
\end{exam}

\begin{claim}\label{claim-example 1}
The Bayesian game in Example~\ref{exam cond independence} satisfies DCPI only when $J \ge 2$.
\end{claim}


\section{The existence result}
\label{sec:exist}


\subsection{Existence of pure-strategy equilibria}
\label{subsec:exist}

In this section, suppose that $u_i(\cdot, t)$ is continuous in $a$ for each $t \in T$ and $i \in I$. We shall prove that under DCPI, pure-strategy equilibria exist in general Bayesian games. Importantly,  this condition is shown to be also necessary for the equilibrium existence result. The proofs are left in Appendix.

\begin{thm}
\label{thm:exist}
Every Bayesian game with decomposable coarser payoff-relevant information has a pure-strategy equilibrium.
\end{thm}

In the theorem above, we show that DCPI is sufficient for the equilibrium existence result. Below, we shall state the necessity part. Rather than proving that this condition must be satisfied if a pure-strategy equilibrium exists in every Bayesian game, we shall prove a stronger result: the necessity result holds even when we restrict our attention to a class of very simple games.

To state the necessity part, we repeat the description of the private type spaces for clarity.
\begin{enumerate}

\item The private type space of player~$i \in I$ is $(T_i, \cT_i/\cF_i, \lambda_i)$ with the payoff-relevant information $\cF_i$.
\end{enumerate}

We focus on a special class of games with incomplete information as follows.

\begin{enumerate}
\item[2.] Players have type-irrelevant payoffs in the sense that the payoff function of each player does not depend on the type profile $t$.
\end{enumerate}

Fix an arbitrary infinite compact set $X$. Let $\Gamma_n(X)$ be the collection of all Bayesian games with the player space $I$, the common action set $X$, the private type spaces $\{(T_i, \cT_i/\cF_i, \lambda_i)\}_{i \in I}$, and type-irrelevant payoffs. A Bayesian game belonging to the class $\Gamma_n(X)$ has a simple structure: players' payoffs only depend on the action profile, but not on the types at all. One may wonder whether we need to construct some games with rather complicated information structures in order to prove the necessity part. Working with the class of simple games in $\Gamma_n(X)$ addresses this concern.

\begin{prop}
\label{prop:exist}
Fix any infinite compact set $X$. If every Bayesian game in $\Gamma_n(X)$ has a pure-strategy equilibrium, then the condition of decomposable coarser payoff-relevant information holds.
\end{prop}

To illustrate the DCPI condition and Theorem~\ref{thm:exist}, we consider Bayesian games with interdependent payoffs, where  players' private information are independent conditioned on finitely many states. This framework is commonly adopted in applications. We show that DCPI holds for this class of Bayesian games.

\begin{itemize}
\item For each $i \in I$, player~$i$'s private information space is $(T_i, \cT_i)$.

\item Let $T_0 = \{t_{01}, \ldots, t_{0J}\}$ be the space of unobservable common states that affect the payoffs of all players. The state $t_{0j}$ happens with probability $\tau^j > 0$ for $1 \le j \le J$.

\item Given each $t_{0j} \in T_0$, let $\lambda^j$ be the conditional prior on $(\prod_{i \in I}T_i, \; \otimes_{i \in I} \cT_i)$. The marginal $\lambda_i^j$ of $\lambda^j$ on $(T_i, \cT_i)$ is atomless and $\lambda^j = \otimes_{i \in I}\lambda_i^j$.

\item For each $i$, player~$i$'s payoff function $u_i(a, t_0, t_1, \dots, t_n)$ is $\otimes_{\ell \in I} \cF_{\ell}$-measurable for each $(a, t_0) \in A \times T_0$.

\item The model is viewed as an $n$-player game, in which the space of type profiles is $T = \prod_{1 \le i \le n}T_i$, the common prior is $\hat{\lambda} = \sum_{1 \le j \le J} \tau^j \lambda^j$, and the marginal of $\hat{\lambda}$ is $\hat{\lambda}_{\ell} = \sum_{1 \le j \le J} \tau^j \lambda_{\ell}^j$ for each $\ell \in I$.
\end{itemize}

\begin{coro}\label{coro-interdependence}
If $\cT_{\ell}$ is nowhere equivalent to $\cF_{\ell}$ under $\hat{\lambda}_{\ell}$ for each $\ell \in I$, then the above Bayesian game with interdependent payoff and conditionally independent types satisfies DCPI, and has pure-strategy equilibria.
\end{coro}

When $u_i(a, t_0, \cdot)$ does not depend on $t_{-i}$ for each $i \in I$, the game defined above is reduced to a Bayesian game with private values and conditionally independent types. If we impose the condition that $\cT_{\ell}$ is nowhere equivalent to $\cF_{\ell}$ for each $\ell \in I$, then it follows directly from Corollary~\ref{coro-interdependence} that pure-strategy equilibria exist in Bayesian games with private values and conditionally independent types.

\begin{coro}\label{coro-private}
Suppose that for each $\ell \in I$, $\cT_{\ell}$ is nowhere equivalent to $\cF_{\ell}$ under $\hat{\lambda}_{\ell}$ and $u_{\ell}(a, t_0, \cdot)$ does not depend on $t_{-\ell}$ for each $a \in A$, $t_0 \in T_0$ and $\ell \in I$. Then the Bayesian game with private values and conditionally independent types as defined above satisfies DCPI, and has pure-strategy equilibria.
\end{coro}

\begin{rmk}
\label{rmk:exist}
\rm

A major idea to prove the existence of pure-strategy equilibria in Bayesian games with general action spaces is to assume rich structures on the type spaces; see \cite{KS1999} and \cite{LS2006} for atomless Loeb spaces, \cite{WZ2012} and \cite{KZ2014} for saturated probability spaces, and \cite{HS2014} for the relative diffuseness condition. In this paper, we are able to relax these additional richness conditions. Furthermore, all those papers work with Bayesian games with private values and (conditionally) independent types. Our Corollary~\ref{coro-private} covers the corresponding results in those papers as special cases.
Note that our Theorem~\ref{thm:exist} (and Corollary~\ref{coro-interdependence}) allows for interdependent payoffs and correlated types.
\end{rmk}


\subsection{A Cournot duopoly game}
\label{subsec:application}

Below, we shall present a simple example of Cournot duopoly to illustrate the model and the condition of DCPI.

There are two firms, $I = \{1, 2\}$. The market could be in one of the two possible unknown status: $H$ or $L$, where $H$ represents boom and $L$ represents recession. That is, the set $T_0 = \{H, L\}$ is the common state space, with $t_0 \in T_0$ being unobservable to both firms. Each firm~$i$ will possess a bi-dimensional private type $t_i = (t_{i1}, t_{i2}) \in T_{i1} \times T_{i2}$, where $T_{11} = T_{12} = T_{21} = T_{22} = [0, 1]$ endowed with the Borel $\sigma$-algebra. The first component $t_{i1}$ summarizes firm~$i$'s information about the basic structure of the market, including demand shock, regulation policy, production cost, etc. The second component $t_{i2}$ affects firm~$i$'s belief about the market status $t_0$, though it does not directly enter the payoffs. Based on the private information $(t_{i1}, t_{i2})$, firm~$i$ chooses a quantity $a_i \in [0, \bar{a}_i]$ to produce. For each firm~$i$, the payoff function $u_i$ is bounded and continuous, and depends on $(a_1, a_2)$ and $(t_0, t_{11}, t_{21})$. The information structure is described as follows.

\begin{itemize}
\item The common states $H$ and $L$ are drawn with probabilities $\frac{1}{3}$ and $\frac{2}{3}$, respectively.

\item The first components of firms' types, $t_{11}$ and $t_{21}$, are drawn according to the uniform distribution $\eta$ on $T_{11} = T_{21} = [0, 1]$, and are independent of each other and also the other components $t_0, t_{12}, t_{22}$.

\item If $t_0 = H$, then a pair $(t_{12}, t_{22}) \in T_{12} \times T_{22} = [0, 1] \times [0, 1]$ will be drawn according to the distribution $\tilde{\zeta}$, which has the density $(\frac{1}{2} + t_{12}) (\frac{1}{2} + t_{22})$ with respect to $\eta \otimes \eta$.

\item If $t_0 = L$, then a pair $(t_{12}, t_{22}) \in T_{12} \times T_{22} = [0, 1] \times [0, 1]$ will be drawn according to the distribution $\hat{\zeta}$, which has the density $4t_{12} t_{22}$ with respect to $\eta \otimes \eta$.
\end{itemize}

In the example above, the first component $t_{i1}$, together with the common state $t_0$, is payoff relevant to both firms. The second component $t_{i2}$ does not enter the payoffs directly. However, since $t_0$ is not observable, $t_{i2}$ is a signal that firm~$i$ receives about the common state, which further affects firm~$i$'s belief about the opponent's type $t_{j2}$, $j \neq i$. Let $\tilde{\lambda} = \eta \otimes \eta \otimes \tilde{\zeta}$,\footnote{Here we abuse the notion by using $\eta \otimes \eta \otimes \tilde{\zeta}$ to denote a probability measure on $T_{11} \times T_{12} \times T_{21} \times T_{22}$ such that $\eta \otimes \eta \otimes \tilde{\zeta}(B_1 \times B_2 \times D_1 \times D_2) = \eta(B_1) \cdot \eta(D_1) \cdot \tilde{\zeta}(B_2 \times D_2)$ for any Borel subsets $B_1$, $B_2$, $D_1$, and $D_2$ in $[0, 1]$. The notion $\eta \otimes \eta \otimes \hat{\zeta}$ is similar.} and $\hat{\lambda} = \eta \otimes \eta \otimes \hat{\zeta}$. In this Cournot duopoly game, the space of type profiles is $T = T_1 \times T_2$ and the common prior is $\lambda = \frac{1}{3}\tilde{\lambda} + \frac{2}{3}\hat{\lambda}$.

The following claim follows from Corollary~\ref{coro-interdependence}.

\begin{claim}
\label{claim:cournot}
The Cournot game above satisfies the condition of decomposable coarser payoff-relevant information, and possesses a pure-strategy equilibrium.
\end{claim}


\section{Purification}
\label{sec:purification}

In this section, we consider the purification method for general Bayesian games. In Section~\ref{subsec:purification}, we establish a general purification principle relating behavioral strategies to pure strategies, which preserves the same expected payoffs and action distributions, and thus the equilibrium property. It is shown that DCPI is both necessary and sufficient for the purification in Bayesian games. Importantly, the purification result does not impose the continuity condition on payoffs. As a result, we are able to apply it to discontinuous Bayesian games and obtain a new equilibrium existence result in pure strategies.


\subsection{A purification result}
\label{subsec:purification}

In this section, we shall define the notion of conditional purification, and prove that DCPI is sufficient and necessary for the existence of conditional purifications.

\begin{defn}
\label{defn:purification}
\rm
Let $f = (f_1, f_2, \ldots, f_n)$ and $g = (g_1, g_2, \ldots, g_n)$ be two behavioral-strategy profiles.
\begin{enumerate}
\item The strategy profiles $f$ and $g$ are said to be payoff equivalent for player~$i$ if $U_i(f) = U_i(g)$, and $U_i(h_i, f_{-i}) = U_i(h_i, g_{-i})$ for any given behavioral strategy $h_i$ of player $i$.

\item The strategy profiles $f$ and $g$ are said to be conditional distribution equivalent for player $i$ if for any $D \in \cF_i$ and Borel subset $B \subseteq X$, $\int_{D} f_i(t_i; B) \lambda_i(\dif t_i) = \int_D g_i(t_i; B) \lambda_i(\dif t_i)$.

\item When $f_i$ is a pure-strategy for player $i$, $f$ and $g$ are said to be belief consistent for player~$i$ if $f_i(t_i)\in \supp g_i(t_i)$ for $\lambda_i$-almost all $t_i \in T_i$, where $\supp g_i(t_i)$ is the support of the probability measure $g_i(t_i)$.
\end{enumerate}
A pure-strategy profile $f$ is said to be a \emph{conditional purification} of a behavioral-strategy profile $g$ if $f$ and $g$ are payoff equivalent, conditional distribution equivalent, and belief consistent for every player.
\end{defn}

The notion of conditional purification requires that from each player's perspective, a behavioral-strategy profile and its purification are the same when the player evaluates the payoff and action distribution conditioned on any non-trivial payoff-relevant event. It is clear that if a behavioral-strategy profile $g$ is an equilibrium, then its conditional purification $f$ is a pure-strategy equilibrium.

We focus on Bayesian games with private type spaces $\{(T_i, \cT_i/\cF_i, \lambda_i)\}_{i \in I}$. The following result characterizes the existence of conditional purifications for these games via DCPI.

\begin{thm}
\label{thm:purification}
In Bayesian games with private type spaces $\{(T_i, \cT_i/\cF_i, \lambda_i)\}_{i \in I}$, every behavioral-strategy profile $g$ possesses a conditional purification $f$ if and only if the condition of decomposable coarser payoff-relevant information holds.
\end{thm}


\subsection{Discontinuous Bayesian games}
\label{subsec:discontinuous}

Payoff discontinuity in actions is a natural feature in various economic applications of Bayesian games; \textit{e.g.}, auctions, contests, price competitions, etc.\footnote{There is a large literature studying discontinuous games in the last two decades. For some recent developments, see, for example, \cite{Prokopovych2011}, \cite{CM2019} and \cite{PY2021}. The literature is too vast to be discussed in the context of this paper, we refer the interested readers to the recent survey by \cite{Reny2020}.} \cite{CM2018} and \cite{HY2016} proved the existence of equilibria in behavioral strategies for Bayesian games with discontinuous payoffs. In this section, we shall demonstrate the usefulness of the purification result via discontinuous Bayesian games. In particular, by verifying DCPI in these games, we obtain a new existence result on pure-strategy equilibria through the purification principle from Theorem~\ref{thm:purification}. As an illustrative example, we present a model of common-value all-pay auctions with general value functions and tie-breaking rules, and obtain a new existence result for pure-strategy equilibria.

To guarantee the existence of behavioral-strategy equilibria, \cite{CM2018} and \cite{HY2016} proposed the conditions of  ``uniform payoff security'' and ``random disjoint payoff matching,'' respectively.\footnote{\cite{CM2018} also considered the equilibrium existence in distributional strategies under the same condition.} Both conditions are to ensure that the ex ante payoff $U_i$ is payoff secure for each player~$i \in I$, which is the key condition in \cite{Reny1999}. Thus, one can readily apply \cite{Reny1999}'s result to conclude the equilibrium existence. To be concrete, the conditions are stated in the following.

\begin{defn}[Uniform payoff security]
\label{defn:uniform}
\rm
The Bayesian game is said to be \emph{uniformly payoff secure} if for any $\epsilon > 0$, each $i \in I$, and each pure strategy $f_i$, there exists another pure strategy $f'_i$ such that for all $(t, a_{-i})$, there exists a neighborhood $O_{a_{-i}}$ of $a_{-i}$ such that for all $y_{-i} \in O_{a_{-i}}$,
$$ u_i \bigl( t, f'_i(t_i), y_{-i} \bigr)  -  u_i \bigl( t, f_i(t_i), a_{-i} \bigr)  >  - \epsilon. $$
\end{defn}

\begin{defn}[Random disjoint payoff matching]
\label{defn:DPM}
\rm
Consider the points at which a player's payoff function is discontinuous in other players' strategies. Let $D_i \colon T_i\times A_i \to T_{-i} \times A_{-i}$ be defined by
$$ D_i(t_i, a_i)  =  \{ (t_{-i}, a_{-i}) \in T_{-i} \times A_{-i}  \mid  u_i(a_i, \cdot, t_i, t_{-i}) \mbox{ is discontinuous in } a_{-i} \}. $$
Suppose that $D_i$ has a $\cB(A_i) \otimes \cT_i$-measurable graph for each $i \in I$. Given a pure strategy $f_i$ of player~$i$, denote $D_i^{f_i}(t_i) = D_i \bigl( t_i, f_i(t_i) \bigr)$.

A Bayesian game $G$ is said to satisfy the condition of \emph{random disjoint payoff matching} if for each player $i \in I$ and for any pure strategy $f_i$, there exists a sequence of pure-strategy deviations $\{h_i^k\}_{k=1}^\infty$ such that
\begin{enumerate}
\item for $\lambda$-almost all $t = (t_i, t_{-i}) \in T$ and for all $a_{-i} \in A_{-i}$,
    $$ \liminf_{k \to \infty} u_i \bigl( f_i^k(t_i), a_{-i}, t_i, t_{-i} \bigr)  \ge  u_i \bigl( f_i(t_i), a_{-i}, t_i, t_{-i} \bigr); $$
\item $\limsup\limits_{k \to \infty}  D_i \bigl( t_i, f^k_i(t_i) \bigr)  =  \emptyset$ for each $i \in I$ and for $\lambda_i$-almost all $t_i \in T_i$.\footnote{For a sequence of sets $\{X_k\}$, $\limsup\limits_{k \to \infty} X_k = \cap_{k=1}^\infty \cup_{j=k}^\infty X_j$ and $\liminf\limits_{k\to\infty} X_k = \cup_{k=1}^\infty \cap_{j=k}^\infty X_j$.}
\end{enumerate}
\end{defn}

The following lemma summarizes Theorem~1 in \cite{CM2018} and Theorem~2 in \cite{HY2016}.

\begin{lem}
\label{lem:secure}
Suppose that the aggregate payoff $\sum_{i \in I} u_i(\cdot, t) \colon A \to \bR$ is upper semicontinuous for each $t \in T$. If any of the following conditions holds,
\begin{enumerate}
\item the uniform payoff security,
\item the random disjoint payoff matching,
\end{enumerate}
then the game has a behavioral-strategy equilibrium.
\end{lem}

It is clear that based on Theorem~\ref{thm:purification} and Lemma~\ref{lem:secure} above, one can obtain the existence of pure-strategy equilibria by assuming the DCPI condition in discontinuous Bayesian games.

\begin{prop}
\label{prop:discontinuous}
Suppose that the aggregate payoff $\sum_{i \in I} u_i(\cdot, t) \colon A \to \bR$ is upper semicontinuous for each $t \in T$, and one of the following conditions holds,
\begin{enumerate}
\item the uniform payoff security,

\item the random disjoint payoff matching.
\end{enumerate}
If the discontinuous game has decomposable coarser payoff-relevant information, then it possesses a pure-strategy equilibrium.
\end{prop}

Below, we provide an example of a common-value all-pay auction with general value functions and tie-breaking rules. \citet[Section~6.1]{CM2018} studied an all-pay auction with general value functions and standard tie-breaking rules. \citet[Section~4]{HY2016} presented an all-pay auction with quasi-linear payoffs and general tie-breaking rules. The following example cannot be covered by the all-pay auctions as considered in those two papers.

\begin{exam}
\label{exam:allpay}
\rm
Suppose that $n \ge 2$ bidders compete for an object. Let $I = \{1, 2, \ldots, n\}$. For each $i \in I$, bidder~$i$ observes a private type $t_i = (t_{i1}, t_{i2}) \in T_i$ and submits a bid $a_i$ from the bid space $A_i$, where $T_i = T_{i1} \times T_{i2}$ is a compact rectangle in $\bR^2$ endowed with the Borel $\sigma$-algebra $\cT_i = \cB(T_i)$, and $A_i = [0, \bar{a}] \subseteq \bR_{+}$ with $\bar{a} > 0$. An unobservable common type $t_0^j$ is drawn from a finite type space $T_0 = \{t_0^1, t_0^2, \ldots, t_0^J\}$ with probability $\tau^j > 0$, and $\sum_{j=1}^J \tau^j = 1$. Given $t_0^j$, the types $(t_1, t_2, \ldots, t_n)$ is drawn from $\prod_{i \in I} T_i$ independently according to a Borel probability measure $\otimes_{i \in I}\lambda^j_i$. For each $i \in I$ and for each $j = 1, 2, \ldots, J$, the density of $\lambda^j_i$ is $q_i^j > 0$ (with respect to the Lebesgue measure on $T_i$). The common prior is $\lambda = \sum_{j=1}^J \tau^j \otimes_{i \in I} \lambda^j_i$. Let $T = \prod_{i \in I} T_i$ and $A = \prod_{i \in I}A_i$. For convenience, denote $\tilde{T} = T_0 \times T$.

Given $\tilde{t} = \bigl( t_0, t_1, \ldots, t_n \bigr)$ and $a = (a_1, a_2, \ldots, a_n)$, let
$$ v_i(\tilde{t}, a) = \begin{cases}
\psi_1(\tilde{t}, a) + \varphi_i(\tilde{t}, a), & \text{if bidder $i$ is the unique winning bidder},    \\
\psi_2(\tilde{t}, a) + \varphi_i(\tilde{t}, a), & \text{if bidder $i$ is a losing bidder}.
\end{cases}
$$
Intuitively, one can interpret $\psi_1(\tilde{t}, a)$ as the common value for winning the object, $\psi_2(\tilde{t}, a)$ as the common outside option when losing it, and $\varphi_i(\tilde{t}, a)$ as the cost of bidding.
The bidder with the highest bid wins the object. Ties are broken as follows: if $ a_i = \max_{k \in I} a_k$,
$$ v_i(\tilde{t}, a)  =  \sigma_i(a) \psi_1(\tilde{t}, a)  +  \bigl( 1-\sigma_i(a) \bigr)  \psi_2(\tilde{t}, a)  +  \varphi_i(\tilde{t}, a), $$
where $\sigma_i(a_1, a_2, \ldots, a_n) = \frac{\xi_i(a_1, a_2, \ldots, a_n)}{\sum\limits_{\ell \in I \colon a_\ell = \max_{k \in I} a_k} \xi_\ell(a_1, a_2, \ldots, a_n)}$ and $\xi = (\xi_1, \xi_2, \ldots, \xi_n) \colon [0, \bar{a}]^n \to (0, 1]^n$ is a continuous function measuring the relative importance of each bidder's position when breaking a tie.\footnote{If $\xi_i \equiv 1$ for each $i$, then a tie is broken via the standard equal proportion rule.} We assume that $\xi_i$ is increasing in $a_i$ and decreasing in $a_{-i}$; that is, when bidder~$i$ increases his bid, he is more likely to win, while the others are more likely to lose.

For any $\tilde{t} = \bigl( t_0, t_1, \ldots, t_n \bigr)$, we further assume that $\psi_1(\tilde{t}, a) = \psi_1(\tilde{t}^1, a)$, $\psi_2(\tilde{t}, a) = \psi_2(\tilde{t}^1, a)$, and $\varphi_i(\tilde{t}, a) = \varphi_i(\tilde{t}^1, a)$ for each $i \in I$, where $\tilde{t}^1 = (t_0, t_{11}, t_{21}, \ldots, t_{n1})$. Thus, the first coordinate $t_{i1}$ of $t_i$ directly affects the payoffs of all the players. Although the second coordinate $t_{i2}$ of $t_i$ is payoff irrelevant, it influences player~$i$'s belief about other players' types.

We assume that
\begin{enumerate}
\item $\psi_1$, $\psi_2$, and $\{\varphi_i\}_{i \in I}$ are continuous on $\tilde{T} \times A$;

\item winning the object is always better than losing it, that is, for any $(t_0, t_{11}, \ldots, t_{n1}, a)$,
$$ \psi_1(t_0, t_{11}, \ldots, t_{n1}, a) \ge \psi_2(t_0, t_{11}, \ldots, t_{n1}, a). $$
\end{enumerate}
Define a transition probability $\kappa$ from $T = \prod_{i \in I} T_i$ to $\cM(T_0)$ such that for $j = 1, 2, \ldots, J$,
$$ \kappa(\{t^j_0\} \mid t)  =  \frac{\tau^j \prod_{\ell \in I} q_\ell^j(t_\ell)}{\sum_{r=1}^J \tau^r \prod_{\ell \in I} q_\ell^r(t_\ell)}, $$
which is the conditional probability of the common type $t^j_0$ given the realized type profile $t \in T$. When the type profile is $t = (t_1, t_2, \ldots, t_n)$ and the bidding profile is $a = (a_1, a_2, \ldots, a_n)$, bidder~$i$'s payoff is
$$ u_i(t_1, t_2, \ldots, t_n, a_1, a_2, \ldots, a_n)  =  \sum_{j=1}^J  v_i(t^j_0, t_1, t_2, \ldots, t_n, a_1, a_2, \ldots, a_n) \kappa(\{t^j_0\} \mid t). $$
\end{exam}

In the all-pay auction above, we consider the environment in which bidders may have interdependent value functions, outside options, and general tie-breaking rules. We show that the condition of uniform payoff security is satisfied and the aggregate payoff is upper semicontinuous. Then there exists a behavioral-strategy equilibrium. We  conclude the existence of a pure-strategy equilibrium by applying the purification result.

\begin{claim}
\label{claim:allpay}
A pure-strategy equilibrium exists in the above all-pay auction with general value functions and tie-breaking rules.
\end{claim}

\section{Appendix}
\label{sec:appendix}

\subsection{Technical preparations}
\label{subsec:rcd}

In this section, we present several results as the mathematical preparations for the proofs of Theorems~\ref{thm:exist} and \ref{thm:purification}, and Proposition~\ref{prop:exist}.

Let $(T, \hat\cT)$ be a measurable space, $\hat\cF$ a countably-generated sub-$\sigma$-algebra of $\hat\cT$, and $\mu_j$ an atomless finite positive measure on $(T, \hat\cT)$ for $j = 1, 2, \ldots, J$. Denote $\mu = (\mu_1, \mu_2, \ldots, \mu_J)$. Suppose that each $\mu_j$ is absolutely continuous with respect to some atomless probability measure $\mu_0$. Let $X$ be a Polish space (complete metrizable topological space), $\cB(X)$ the Borel $\sigma$-algebra of $X$, and $\cM(X)$ the space of all Borel probability measures on $X$ endowed with the topology of weak convergence.

An $\hat\cF$-measurable \emph{transition probability} from $T$ to $X$ is a mapping $\phi \colon T \to \cM(X)$ such that for every $B \in \cB(X)$, the mapping $\phi(\cdot; B) \colon t \mapsto \phi(t; B)$ is $\cF$-measurable, where $\phi(t; B)$ is the value of the probability measure $\phi(t)$ on the Borel set $B \subseteq A$. For each $j = 1, 2, \ldots, J$, we use $\cR^{(\hat\cF, \mu_j)}(X)$, or $\cR^{(\hat\cF, \mu_j)}$ when it is clear, to denote the set of all $\hat\cF$-measurable transition probabilities from $T$ to $X$ under $\mu_j$. The set $\cR^{(\hat\cF, \mu_j)}$ is endowed with the following weak topology.

A sequence $\{\phi_n\}_{n=1}^\infty$ in $\cR^{(\hat\cF, \mu_j)}$ is said to \emph{weakly converge} to $\phi$ in $\cR^{(\hat\cF, \mu_j)}$, if for every bounded Carath\'eodory function $c \colon T \times X \to \bR$,\footnote{A function $c \colon T \times X \to \bR$ is a Carath\'eodory function if $c(\cdot, x)$ is $\hat\cF$-measurable for each $x \in X$ and $c(t, \cdot)$ is continuous for each $t \in T$.}
$$ \lim_{n \to \infty}  \int_T  \left[ \int_X c(t, x) \phi_n(t; \dif x) \right]  \mu_j(\dif t)  =  \int_T  \left[ \int_X c(t, x) \phi(t; \dif x) \right]  \mu_j(\dif t). $$
The weak topology on $\cR^{(\hat\cF, \mu_j)}$ is defined as the weakest topology for which the functional
$$ \phi  \mapsto  \int_T  \left[ \int_X c(t, x) \phi(t; \dif x) \right]  \mu_j(\dif t) $$
is continuous for every bounded Carath\'eodory function $c \colon T \times X \to \bR$.

Denote $\cR^{(\hat\cF, \mu)} = \prod_{j=1}^J \cR^{(\hat\cF, \mu_j)}$. Similarly, one can define $\cR^{(\hat\cT, \mu_j)}$ for each $k = 1, 2, \ldots, J$ and $\cR^{(\hat\cT, \mu)} = \prod_{j=1}^J \cR^{(\hat\cT, \mu_j)}$.

\bigskip

Next, we review the notion of regular conditional distribution. Let $f$ be a $\hat\cT$-measurable mapping from $T$ to $X$. A mapping $\mu_j^{f \mid \hat\cF} \colon T \times \cB(X) \to [0, 1]$ is said to be a \emph{regular conditional distribution} of $f$ given $\hat\cF$ under $\mu_j$, if
\begin{enumerate}
\item for $\mu_j$-almost all $t \in T$, $\mu_j^{f \mid \hat\cF}(t, \cdot)$ is a probability measure on $X$;
\item for each Borel subset $B \subseteq X$, $\mu_j^{f \mid \hat\cF}(\cdot, B)$ is a version of $\bE[1_B(f) \mid \hat\cF]$, where $\bE[1_B(f) \mid \hat\cF]$ is the conditional expectation of the indicator function $1_B(f)$ given $\hat\cF$ under $\mu_j$.
\end{enumerate}

Let $F$ be a correspondence from $T$ to $X$. We use
$$ \cR^{(\hat\cT, \hat\cF, \mu)}_F  =  \left\{ \mu^{f \mid \hat\cF} = \bigl( \mu_1^{f \mid \hat\cF}, \mu_2^{f \mid \hat\cF}, \ldots, \mu_J^{f \mid \hat\cF} \bigr)  \;\Big\vert\;  f \text{ is a $\hat\cT$-measurable selection of $F$} \right\} $$
to denote the set of all regular conditional distributions induced by $\hat\cT$-measurable selections of $F$ conditional on $\hat\cF$ under the vector measure $\mu = (\mu_1, \mu_2, \ldots, \mu_J)$.

The following result is a direct corollary of \citet[Theorem~2]{HS2021}, which presents several desirable properties for the conditional expectation of Banach-valued correspondences.\footnote{The theory of correspondences has been well developed based on Loeb spaces and saturated probability spaces; see, for example, \cite{Sun1996}, \cite{Podczeck2008} and \cite{SY2008}. These results exclude many widely-adopted probability spaces, including the Euclidean spaces. Lemma~\ref{lem:rcd} does not have this restriction. }

\begin{lem}
\label{lem:rcd}
If $\hat\cT$ is nowhere equivalent to $\hat\cF$ under $\mu_0$, then we have the following results.
\begin{enumerate}
\item For any compact-valued $\hat\cF$-measurable correspondence $F$ from $T$ to $X$, $\cR^{(\hat\cT, \hat\cF, \mu)}_F$ is convex and weakly compact.

\item Let $F$ be a compact-valued $\hat\cF$-measurable correspondence from $T$ to $X$, $Z$ a metric space, and $G$ a closed-valued correspondence from $T \times Z$ to $X$ such that
	\begin{itemize}
	\item for each $(t, z) \in T \times Z$, $G(t, z) \subseteq F(t)$;
	\item for each $z \in Z$, $G(\cdot, z)$ is $\cF$-measurable from $T$ to $X$;
	\item for each $t \in T$, $G(t, \cdot)$ is upper-hemicontinuous from $Z$ to $X$.
	\end{itemize}
	Then $H(z) = \cR^{(\hat\cT, \hat\cF, \mu)}_{G(t, z)}$ is upper-hemicontinuous in $z$ from $Z$ to $\cR^{(\hat\cF, \mu)} = \prod_{j=1}^J \cR^{(\hat\cF, \mu_j)}$.
	
\item For any $g = (g_1, g_2, \ldots, g_J) \in \cR^{(\hat\cF, \mu)} = \prod_{j=1}^J \cR^{(\hat\cF, \mu_j)}$, there exists a $\hat\cT$-measurable mapping $f \colon T \to X$ such that $g = \mu^{f \mid \hat\cF}$, that is, for each $j = 1, 2, \ldots, J$, $g_j = \mu_j^{f \mid \hat\cF}$.

\end{enumerate}
\end{lem}

\subsection{Proofs of Theorem~\ref{thm:exist} and Corollary~\ref{coro-interdependence}}
\label{subsec:proof:exist}

The following lemma will be useful for proving Theorem~\ref{thm:exist}.

\begin{lem}
\label{lem:rcd-2variables}
Let $(T, \cT, \mu)$ be a probability space, $\cF$ a sub-$\sigma$-algebra of $\cT$, $h$ a $\cT$-measurable mapping from $T$ to a Polish space $X$ with the Borel $\sigma$-algebra $\cB$, and $\mu^{h \mid \cF}$ a regular conditional distribution of $h$ given $\cF$ under $\mu$. Let $\psi$ be a $\cB \otimes \cF$-measurable function from $X \times T$ to $\bR$ which is integrably bounded; that is, $|\psi(x, t)| \le \varphi(t)$ for all $(x, t) \in X \times T$, where $\varphi$ is integrable on $(T, \cT, \mu)$. Then, for $\mu$-almost all $t \in T$,
\begin{equation}
\label{eq:rcd-2variables}
\bE \bigl[ \psi \bigl( h(t), t \bigr) \mid \cF \bigr]  =  \int_X \psi(x, t) \mu^{h \mid \cF} (t; \dif x).
\end{equation}
\end{lem}

\begin{proof}
Let $\Psi$ be the class of all nonnegative $\cB \otimes \cF$-measurable functions from $X \times T$ to $\bR$ such that Equation~\eqref{eq:rcd-2variables} holds for $\mu$-almost all $t \in T$.

For any sets $B \in \cB$ and $D \in \cF$, let $\psi(x, t) = \mathbf{1}_{B \times D} (x, t) = \mathbf{1}_B(x) \cdot \mathbf{1}_D(t)$. Since $\mu^{h \mid \cF}$ is a regular conditional distribution of $h$ given $\cF$ under $\mu$, we have that for $\mu$-almost all $t \in T$,
$$ \bE\bigl[\mathbf{1}_B\bigl(h(t)\bigr) \mid \cF\bigr]  =  \mu^{h \mid \cF} (t; B) = \int_X \mathbf{1}_B(x) \mu^{h \mid \cF} (t; \dif x). $$
Since $D$ is $\cF$-measurable, for $\mu$-almost all $t \in T$,
$$ \bE\bigl[\mathbf{1}_B\bigl(h(t)\bigr) \mathbf{1}_D(t) \mid \cF\bigr]  =  \mathbf{1}_D(t) \cdot \bE\bigl[\mathbf{1}_B\bigl(h(t)\bigr) \mid \cF\bigr]  =  \int_X \mathbf{1}_B(x) \mathbf{1}_D(t) \mu^{h \mid \cF} (t; \dif x), $$
which implies that Equation~\eqref{eq:rcd-2variables} holds for $\psi = \mathbf{1}_{B \times D}$. Hence, $\mathbf{1}_{B \times D} \in \Psi$.

By the properties of conditional expectation, it is obvious that $\Psi$ is a $\lambda$-system in the sense that (1) the constant function $\mathbf{1}$ is in $\Psi$; (2) for any nonnegative real numbers $\alpha_1, \alpha_2$, any $\psi_1, \psi_2 \in \Psi$, $\alpha_1 \psi_1 + \alpha_2 \psi_2 \in \Psi$; (3) for any increasing sequence of functions $\{\psi_k\}_{k=1}^\infty$ in $\Psi$ with a limit function $\psi$, one has $\psi \in \Psi$. Since the class of measurable rectangles $B \times D$ with $B \in \cB$ and $D \in \cF$ is a $\pi$-class (\textit{i.e.}, closed under the operation of finite intersections), the usual $\pi$-$\lambda$ Theorem implies that Equation~\eqref{eq:rcd-2variables} holds for every nonnegative $\cB \otimes \cF$-integrable function $\psi$ from $X \times T$ to $\bR$; see, for example, Theorem 1.4.3 in \citet[p.16]{CT1997}.

For a $\cB \otimes \cF$-measurable function from $X \times T$ to $\bR$ satisfying the conditions of the lemma, one can consider the positive and negative parts of $\psi$ separately. The rest is clear.
\end{proof}

\bigskip

\begin{proof}[Proof of Theorem~\ref{thm:exist}]
For each $j = 1, 2, \ldots, J$, let $\mu_{ij}$ be a probability measure on $(T_i, \cT_i)$ so that it is absolutely continuous with respect to $\lambda_i$ with the density $\rho^j_i$. Denote $\mu_{i0} = \lambda_i$ and $\mu_i = (\mu_{i0}, \mu_{i1}, \ldots, \mu_{iJ})$.

For each $i \in I$, recall that $\cR^{(\cF_i, \mu_{ij})}$ is the set of $\cF_i$-measurable transition probabilities from $T_i$ to $A_i$ under the probability measure $\mu_{ij}$ for each $j = 0, 1, \ldots, J$, and $\cR^{(\cF_i, \mu_i)} = \prod_{j=0}^J \cR^{(\cF_i, \mu_{ij})}$. Clearly, each $\cR^{(\cF_i, \mu_{ij})}$ is nonempty, convex and weakly compact (under the topology of weak convergence), so is $\cR^{(\cF_i, \mu_i)}$. Let $\cR^{\cF} = \prod_{i \in I} \cR^{(\cF_i, \mu_i)} = \prod_{i \in I} \prod_{j=0}^J \cR^{(\cF_i, \mu_{ij})}$, which is endowed with the product topology.

\medskip

Fix a pure-strategy profile $f = (f_1, f_2, \ldots, f_n)$. For any two distinct players $i$ and $\ell$, $j = 1, 2, \ldots, J$, types $t_{-\ell} \in T_{-\ell}$, and actions $a_{-\ell} \in A_{-\ell}$, the Fubini property implies that the function $w^j_i (a_{-\ell}, \cdot, t_{-\ell}, \cdot)$ is $\cB(A_\ell) \otimes \cF_\ell$-measurable in $(a_\ell, t_\ell)$. Lemma~\ref{lem:rcd-2variables} above implies that for $\mu_{\ell j}$-almost all $t_\ell \in T_\ell$,
\begin{equation}
\label{eq:rcd}
\bE^{\mu_{\ell j}} \Big[ w^j_i \bigl(a_{-\ell}, f_\ell(t_\ell), t_{-\ell}, t_\ell\bigr) \;\Big\vert\; \cF_\ell \Big] = \int_{A_\ell} w^j_i (a_{-\ell}, a_\ell, t_{-\ell}, t_\ell) \mu_{\ell j}^{f_\ell \mid \cF_\ell}(t_\ell; \dif a_\ell),
\end{equation}
where the left hand side is the conditional expectation given $\cF_\ell$ under $\mu_{\ell j}$.

Fix player~$1$. For $j = 1, 2, \ldots, J$, $t_1 \in T_1$, and $a_1 \in A_1$, we have that
\begin{align*}
    & \int_{T_{-1}}  w^j_1  \big( a_1, f_{-1}(t_{-1}), t_1, t_{-1} \big)  \textstyle{\prod\limits_{\ell \ne 1}} \rho^j_\ell(t_\ell)  \lambda_{-1}(\dif t_{-1})  \\
={} & \int_{T_{-(1,2)}} \int_{T_2}  w^j_1  \big( a_1, f_2(t_2), f_{-(1,2)}(t_{-(1,2)}), t_2, t_{-2} \big) \mu_{2j} (\dif t_2) \mu_{-(1,2)j}(\dif t_{-(1,2)})    \\
={} & \int_{T_{-(1,2)}} \int_{T_2}  \bE^{\mu_{2j}} \Big[ w^j_1 \big(a_1, f_2(t_2), f_{-(1,2)}(t_{-(1,2)}), t_2, t_{-2}\big) \;\Big\vert\; \cF_2 \Big] \mu_{2j} (\dif t_2) \mu_{-(1,2)j}(\dif t_{-(1,2)})    \\
={} & \int_{T_{-(1,2)}} \int_{T_2}  \int_{A_2} w^j_1 \big( a_1, a_2, f_{-(1,2)}(t_{-(1,2)}), t_2, t_{-2} \big)  \mu_{2j}^{f_2 \mid \cF_2} (t_2; \dif a_2) \mu_{2j} (\dif t_2) \mu_{-(1,2)j} (\dif t_{-(1,2)}) \\
={} & \cdots \\
={} & \int_{T_{-1}}  \int_{A_{-1}}  w^j_1(a_1, a_{-1}, t_1, t_{-1})  \textstyle{\prod\limits_{\ell \ne 1}} \mu_{\ell j}^{f_\ell \mid \cF_\ell}(t_\ell; \dif a_\ell) \mu_{-1j} (\dif t_{-1}),
\end{align*}
where the subscript $-(1,2)$ denotes all the players except players~$1$ and $2$, and $\mu_{-(1,2)j} = \otimes_{i \ne 1, 2} \mu_{ij}$. The first equality is due to the definition of $\mu_{ij}$, which has density $\rho_i^j$ with respect to $\lambda_i$. The second equality holds by taking the conditional expectation. The third equality follows from Equation~\eqref{eq:rcd}. The last equality follows by repeating these three steps from $T_3$ to $T_n$, which is omitted in the fourth equality. By the definition of the density-weighted payoff, we have that
\begin{align*}
    & \int_{T_{-1}}  u_1 \big( a_1, f_{-1}(t_{-1}), t_1, t_{-1} \big)  q(t_1, t_{-1})  \lambda_{-1}(\dif t_{-1})    \\
={} & \int_{T_{-1}}  w_1 \big( a_1, f_{-1}(t_{-1}), t_1, t_{-1} \big)  \lambda_{-1}(\dif t_{-1})    \\
={} & \int_{T_{-1}}  \sum_{j=1}^J  \biggl[ w^j_1 \big( a_1, f_{-1}(t_{-1}), t_1, t_{-1} \big) \textstyle{\prod\limits_{\ell \in I}} \rho^j_\ell (t_\ell) \biggr] \lambda_{-1} (\dif t_{-1}) \\
={} & \sum_{j=1}^J  \rho^j_1(t_1)  \int_{T_{-1}}  w^j_1 \big( a_1, f_{-1}(t_{-1}), t_1, t_{-1} \big)  \textstyle{\prod\limits_{\ell \ne 1}} \rho^j_\ell (t_\ell) \lambda_{-1} (\dif t_{-1})    \\
={} & \sum_{j=1}^J  \rho^j_1(t_1)  \int_{T_{-1}}  \int_{A_{-1}} w^j_1(a_1, a_{-1}, t_1, t_{-1})  \textstyle{\prod\limits_{\ell \ne 1}} \mu_{\ell j}^{f_\ell \mid \cF_\ell}(t_\ell; \dif a_\ell)  \mu_{-1j}(\dif t_{-1}).
\end{align*}
One can repeat the argument for each player $i \in I$ such that for each $t_i \in T_i$ and $a_i \in A_i$,
\begin{align}
\label{eq:epayoff}
    & \int_{T_{-i}}  u_i \big( a_i, f_{-i}(t_{-i}), t_i, t_{-i} \big)  q(t_i, t_{-i})  \lambda_{-i}(\dif t_{-i})    \nonumber\\
={} & \sum_{j=1}^J  \rho^j_i(t_i)  \int_{T_{-i}}  \int_{A_{-i}}  w^j_i(a_i, a_{-i}, t_i, t_{-i})  \textstyle{\prod\limits_{\ell \ne i}} \mu_{\ell j}^{f_\ell \mid \cF_\ell}(t_\ell; \dif a_\ell)  \mu_{-ij}(\dif t_{-i}).
\end{align}

\medskip

For each $i \in I$, let $F_i$ be a mapping from $T_i \times A_i\times \cR^\cF$ to $\bR$, which is defined as follows:
$$ F_i(t_i, a_i, g_1, g_2, \ldots, g_n)  =  \sum_{j=1}^J  \rho^j_i(t_i)  \int_{T_{-i}}  \int_{A_{-i}}  w^j_i(a_i, a_{-i}, t_i, t_{-i})  \textstyle{\prod\limits_{\ell \ne i}}  g_{\ell j}(t_\ell; \dif a_\ell)  \mu_{-ij}(\dif t_{-i}), $$
where each $g_\ell = (g_{\ell 0}, g_{\ell 1}, \ldots, g_{\ell J}) \in \cR^{(\cF_\ell, \mu_\ell)} = \prod_{j=0}^J \cR^{(\cF_\ell, \mu_{\ell j})}$. It is clear that $F_i$ is $\cF_i$-measurable on $T_i$ and continuous on $A_i \times \cR^\cF$.

For each $i \in I$, we consider the best response correspondence $G_i$ from $T_i \times \cR^\cF$ to $A_i$, which is given by
$$ G_i(t_i, g_1, g_2, \ldots, g_n)  =  \argmax_{a_i \in A_i} F_i(t_i, a_i, g_1, g_2, \ldots, g_n). $$
For each $t_i \in T_i$, since $A_i$ is compact and $F_i$ is continuous on $A_i \times \cR^\cF$, Berge's maximal theorem (see Theorem~17.31 in \cite{AB2006} for example) implies that $G_i(t_i, \cdot)$ is nonempty, compact-valued, and upper-hemicontinuous on $\cR^\cF$. We have already known that for any $a_i \in A_i$ and $(g_1, g_2, \ldots, g_n) \in \cR^\cF$, $F_i(\cdot, a_i, g_1, g_2, \ldots, g_n)$ is $\cF_i$-measurable. Then measurable maximal theorem (see Theorem~18.19 in \cite{AB2006} for example) implies that the correspondence $G_i(\cdot, g_1, g_2, \ldots, g_n)$ admits an $\cF_i$-measurable selection. Thus, $\cR_{G_i(\cdot, g_1, g_2, \ldots, g_n)}^{(\cT_i, \cF_i, \mu_i)}$, the set of regular conditional distributions induced by $\cT_i$-measurable selections of correspondence $G_i(\cdot, g_1, g_2, \ldots, g_n)$ on $\cF_i$ under the vector measure $\mu_i = (\mu_{i0}, \mu_{i1}, \ldots, \mu_{iJ})$, is nonempty. Since $\cT_i$ is nowhere equivalent to $\cF_i$ under $\lambda_i = \mu_{i0}$ and $\mu_{ij}$ is absolutely continuous with respect to $\lambda_i$ for each $j = 1, 2, \ldots, J$, we have that $\cT_i$ is nowhere equivalent to $\cF_i$ under $\mu_{ij}$ for each $j = 0, 1, \ldots, J$; see Lemma~3 in \cite{HS2019}. Lemma~\ref{lem:rcd} then implies that $\cR_{G_i(\cdot, g_1, g_2, \ldots, g_n)}^{(\cT_i, \cF_i, \mu_i)}$ is convex, weakly compact-valued, and weakly upper-hemicontinuous on $\cR^\cF = \prod_{i \in I} \cR^{\cF_i}$.

Consider the correspondence $\Phi$ from $\cR^\cF$ to itself as follows:
$$\Phi(g_1,g_2,\ldots,g_n) = \textstyle{\prod\limits_{i \in I}}  \cR_{G_i(\cdot,g_1,g_2,\ldots,g_n)}^{(\cT_i,\cF_i, \mu_i)}.$$
It is clear that $\Phi$ is nonempty, convex, weakly compact-valued, and upper-hemicontinuous on $\cR^\cF$. By Fan-Glicksberg's fixed-point theorem, there exists a fixed point $(g^*_1, g^*_2, \ldots, g^*_n)$ of $\Phi$. That is, for each $i \in I$, $g_i^* = (g_{i0}^*, g_{i1}^*, \ldots, g_{iJ}^*) \in \cR_{G_i(\cdot, g_1^*, g_2^*, \ldots, g_n^*)}^{(\cT_i, \cF_i, \mu_i)}$. Thus, for each $i \in I$, there exists a $\cT_i$-measurable selection $f_i^*$ of $G_i(\cdot, g_1^*, g_2^*, \ldots, g_n^*)$ such that $g_{ij}^* = \mu_{ij}^{f_i^* \mid \cF_i}$ for each $j = 0, 1, \ldots, J$.

Under the pure-strategy profile $(f_1^*, f_2^*, \ldots, f_n^*)$, the payoff of player $i$ is
\begin{align*}
U_i(f^*)    & = \int_T  u_i \left( f_i^*(t_i), f_{-i}^*(t_{-i}), t_i, t_{-i} \right)  \lambda(\dif t)   \\
            & = \int_{T_i}  \int_{T_{-i}}  u_i \left( f_i^*(t_i), f_{-i}^*(t_{-i}), t_i, t_{-i} \right)  q(t_i, t_{-i})  \lambda_{-i}(\dif t_{-i})  \lambda_i(\dif t_i)  \\
            & = \int_{T_i}  \sum_{j=1}^J  \rho^j_i(t_i)  \int_{T_{-i}}  \int_{A_{-i}}  w^j_i(f_i^*(t_i), a_{-i}, t_i, t_{-i})  \textstyle{\prod\limits_{\ell \ne i}} g_{ij}^* (t_\ell; \dif a_\ell)  \mu_{-ij} (\dif t_{-i})  \lambda_i (\dif t_i)	\\
            & = \int_{T_i} F_i \bigl( t_i, f^*_i(t_i), g_1^*, \ldots, g_n^* \bigr) \lambda_i(\dif t_i).
\end{align*}
The first equality is true due to the definition of $U_i$, and the second equality holds based on the Fubini property. The third equality relies on Equation~\eqref{eq:epayoff}, and the fourth equality holds due to the definition of $F_i$. By the choices of $(f_1^*, f_2^*, \ldots, f_n^*)$, we have that for each $i \in I$, $f_i^*$ maximizes $U_i(f_i, f_{-i}^*)$, and hence $(f_1^*, f_2^*, \ldots, f_n^*)$ is a pure-strategy equilibrium.
\end{proof}

Below, we prove Corollary~\ref{coro-interdependence}.

\begin{proof}[Proof of Corollary~\ref{coro-interdependence}]
We only need to verify that DCPI holds. The equilibrium existence result follows from Theorem~\ref{thm:exist}.

Recall that the marginal $\hat{\lambda}_i$ of $\hat{\lambda}$ on $(T_i, \cT_i)$ is $\sum_{1 \le j \le J} \tau^j \lambda_i^j$. Since $\tau^j > 0$, $\lambda^{j}_i$ is absolutely continuous with respect to $\hat{\lambda}_i$ for each $i \in I$ and $1 \le j \le J$. We assume that the Radon-Nikodym derivative is $q_i^j$. It is clear that $\hat{\lambda}$ is absolutely continuous with respect to $\otimes_{l \in I}\hat{\lambda}_l$ with the Radon-Nikodym derivative $q = \sum_{1 \le j \le J} \tau^j \prod_{l \in I} q_l^j$.
For any $t_{0j} \in T_0$, and $D_l \in \cT_l$ for $1 \le l \le n$,
\begin{align*}
& \quad \hat{\lambda} \left( \{t_{0j}\} \times D_1 \times \cdots \times D_n \right) \\
& = \tau^j \int_{T_1} \cdots \int_{T_n} \prod_{1 \le l \le n} \mathbf{1}_{D_l}(t_l) \lambda^j_n({\rmd t_n}) \cdots \lambda^j_1({\rmd t_1}) \\
& = \int_{T_1} \cdots \int_{T_n} \tau^j \prod_{1 \le l \le n} [ \mathbf{1}_{D_l}(t_l) q_l^j(t_l) ] \hat{\lambda}_n({\rmd t_n}) \cdots \hat{\lambda}_1({\rmd t_1}) \\
& = \int_{T_1} \cdots \int_{T_n} \prod_{1 \le l \le n} \mathbf{1}_{D_l}(t_l) \cdot \frac{\tau^j \prod_{1 \le l \le n} q_l^j(t_l)}{\sum_{1 \le j \le J} \tau^j \prod_{l \in I} q_l^j(t_l)} \hat{\lambda}({\rmd (t_1, \ldots, t_n)}).
\end{align*}
Define a transition probability $\nu$ from $\prod_{1 \le l \le n} T_i$ to $\cM(T_0)$ such that for $1 \le j \le J$,
$$\nu(\{t_{0j}\} | t) = \frac{\tau^j \prod_{1 \le l \le n} q_l^j(t_l)}{\sum_{1 \le j \le J} \tau^j \prod_{l \in I} q_l^j(t_l)},
$$
Define a new payoff function $v_i$ on $A \times \prod_{1 \le l \le n} T_i$ as
$$v_i (a, t) = \sum_{1 \le j \le J} u_i(a, t_{0j}, t) \nu(\{t_{0j}\} | t).
$$
Consider the $n$-player game in which player~$i$ has private information space $(T_i, \cT_i)$, action space $A_i$, and payoff function $v_i$. The common prior is $\hat{\lambda}$. Then the density weighted payoff in this game is
\begin{align*}
v_i (a, t) q(t)
& = \left(\sum_{1 \le j \le J} u_i(a, t_{0j}, t) \cdot \nu(\{t_{0j}\} | t) \right) \cdot \left( \sum_{1 \le j \le J} \tau^j \prod_{l \in I} q_l^j(t_l) \right) \\
& = \left(\sum_{1 \le j \le J} u_i(a, t_{0j}, t) \cdot \frac{\tau^j \prod_{1 \le l \le n} q_l^j(t_l)}{\sum_{1 \le j \le J} \tau^j \prod_{l \in I} q_l^j(t_l)} \right) \cdot \left( \sum_{1 \le j \le J} \tau^j \prod_{l \in I} q_l^j(t_l) \right) \\
& = \sum_{1 \le j \le J} u_i(a, t_{0j}, t) \cdot \tau^j \prod_{1 \le l \le n} q_l^j(t_l).
\end{align*}
Let $w^j_l(a, t) = \tau^j \cdot u_l(a, t_{0j}, t)$, and $\rho_l^j = q_l^j$ for $1 \le j \le J$ and $1 \le l \le n$. It is clear that the DCPI condition is satisfied.
\end{proof}

\subsection{Proof of Proposition~\ref{prop:exist}}
\label{subsec:proof:exist-nece}

To prove Proposition~\ref{prop:exist}, we consider the following auxiliary games.

Let $d$ be a metric on the compact metric space $X$. For a fixed integer $m \ge 2$, pick $m$ distinct elements in $X$, which are denoted as $a_1, a_2, \ldots, a_m$. Choose a positive real number $r < 1$ such that the closed balls $\bar{B}(a_k, r) = \{ a \in X \mid d(a_k, a) \le r \}$\footnote{We use $B(a, r)$ to denote the open ball with the center $a$ and the radius $r$, $\bar{B}(a, r)$ to denote the closed ball with the center $a$ and the radius $r$, and $B^c$ to denote the complement of a set $B$.} are disjoint ($k = 1, 2, \ldots, m$). By Urysohn's Lemma (see Lemma~2.46 in \cite{AB2006}) and the property that every closed set is a $G_\delta$ set, there exist continuous functions $\{ \beta_1, \beta_2, \ldots, \beta_m, \gamma \}$ from $X$ to $[0, 1]$ satisfying the following properties:
\begin{itemize}
\item for each $k = 1, 2, \ldots, m$, $\beta_k(a) = 1$ for $a \in \bar{B}(a_k, \frac{r}{2})$, $\beta_k(a) = 0$ for $a \in B(a_k, r)^c$, and $\beta_k(a) \in (0,1)$ for $a \in B(a_k, r) \cap \bar{B}(a_k, \frac{r}{2})^c$;
\item $\gamma(a) = 0$ for $a \in \cup_{k=1}^m \bar{B}(a_k, \frac{r}{2})$ and $\gamma(a) = -5$ for $\left( \cup_{k=1}^m B(a_k, r) \right)^c$.
\end{itemize}

\paragraph{Step 1}

We construct a 2-player Bayesian games $G_2$ as follows. The common action space for the two players is $X$. For each $i = 1, 2$, player $i$ has a private type space $L_i = [0, 1]$ with Borel $\sigma$-algebra. The information structure $\tau$ on $L_1 \times L_2$ is the uniform distribution on the triangle $\{ (l_1, l_2) \mid 0 \le l_1 \le l_2 \le 1 \}$. Given the type profile $(l_1, l_2) \in L_1 \times L_2$, when player 1 chooses action $s_1$ and player 2 chooses action $s_2$, their type-irrelevant payoffs are given by
\begin{align*}
u_1(s_1,s_2, l_1,l_2) = {}& \sum_{k=1}^m \beta_k(s_1) \cdot \beta_k(s_2) \cdot \big( 3-d(s_1, a_k) \big) + \sum_{k=1}^m \beta_k(s_1) \cdot \beta_{k+1}(s_2) \cdot \big( 1-d(s_1, a_k) \big) \\
                            & + \sum_{k=1}^m \sum_{\ell \ne k, k+1} \beta_k(s_1) \cdot \beta_\ell(s_2) \cdot \big( 2-d(s_1, a_k) \big) + \gamma(s_1) - 2, \\
u_2(s_1,s_2, l_1,l_2) = {}& \sum_{k=1}^m \beta_k(s_1) \cdot \beta_{k+1}(s_2) \cdot \big( 3-d(s_2, a_{k+1}) \big) + \sum_{k=1}^m \beta_k(s_1) \cdot \beta_k(s_2) \cdot \big( 1-d(s_2, a_k) \big) \\
                            & + \sum_{k=1}^m \sum_{\ell \ne k, k+1} \beta_k(s_1) \cdot \beta_\ell(s_2) \cdot \big( 2-d(s_2, a_k) \big) + \gamma(s_2) - 2,
\end{align*}
where we adopt the convention $\beta_{m+1} = \beta_1$ and $a_{m+1} = a_1$.

In the following, we will show that only the actions in $\{a_1, a_2, \ldots, a_m\}$ can be chosen with positive probabilities in an equilibrium for both players.

\bigskip

For any $(l_1, l_2) \in L_1 \times L_2$ and any $s_2 \in X$, we consider the player~1' payoff when she chooses various actions.

When player 1 chooses an action $x \in \left( \cup_{k=1}^m B(a_k, r) \right)^c$, then each $\beta_k(x) = 0$ and $\gamma(x) = -5$, and hence her payoff is $u_1(x, s_2, l_1, l_2) = \gamma(x) - 2 = -7$.

When player 1 chooses the action $a_1$, her payoff is
$$u_1(a_1, s_2, l_1, l_2) = \begin{cases}
-2,								& \text{if } s_2 \in \left( \cup_{k=1}^m B(a_k,r) \right)^c; \\
3 \cdot \beta_1(s_2) - 2,		& \text{if } s_2 \in B(a_1,r); \\
\beta_2(s_2) - 2,				& \text{if } s_2 \in B(a_2,r); \\
2 \cdot \beta_\ell(s_2) - 2,	& \text{if } s_2 \in B(a_\ell,r) \text{ for some $\ell \ne 1,2$}.
\end{cases}$$

Since for each $k = 1, 2, \ldots, m$, $0 \le \beta_k(a) \le 1$ for any $a \in X$, the action $x \in \left( \cup_{k=1}^m B(a_k,r) \right)^c$ is strictly dominated by $a_1$ for player 1. Therefore, every action in $\left( \cup_{k=1}^m B(a_k,r) \right)^c$ will not be chosen with positive probability in an equilibrium for player 1. That is, only actions in $X' = \cup_{k=1}^m B(a_k,r)$ can be chosen with positive probabilities in an equilibrium for player 1.

Similarly, only the actions in $X' = \cup_{k=1}^m B(a_k,r)$ can be chosen with positive probabilities in an equilibrium for player 2.

\bigskip

For any $(l_1, l_2) \in L_1 \times L_2$ and any $s_2 \in X'$, we also consider the player~1' payoff when she chooses various actions in $X'$.

When player 1 chooses an action $x' \in B(a_1,r) \setminus \{a_1\}$, her payoff is
$$\begin{cases}
\beta_1(x') \cdot \beta_1(s_2) \cdot \big(3-d(x', a_1)\big) + \gamma(x') - 2,		& \text{if } s_2 \in B(a_1,r); \\
\beta_1(x') \cdot \beta_2(s_2) \cdot \big(1-d(x', a_1)\big) + \gamma(x') - 2,		& \text{if } s_2 \in B(a_2,r); \\
\beta_1(x') \cdot \beta_\ell(s_2) \cdot \big(2-d(x', a_1)\big) + \gamma(x') - 2,	& \text{if } s_2 \in B(a_\ell,r) \text{ for some $\ell \ne 1, 2$}.
\end{cases}$$

On the other hand, when player 1 chooses the action $a_1$, her payoff is
$$\begin{cases}
3 \cdot \beta_1(s_2) - 2,		& \text{if } s_2 \in B(a_1,r);  \\
\beta_2(s_2) - 2,				& \text{if } s_2 \in B(a_2,r); \\
2 \cdot \beta_\ell(s_2) - 2,	& \text{if } s_2 \in B(a_\ell,r) \text{ for some $\ell \ne 1, 2$}.
\end{cases}$$

Since for each $k = 1, 2, \ldots, m$, $0 \le \beta_k(a) \le 1$ for any $a \in X$, the action $x' \in B(a_1,r) \setminus \{a_1\}$ is strictly dominated by $a_1$ for player 1. For $k = 2, 3, \ldots, m$, similar arguments show that every action in $B(a_k,r) \setminus \{a_k\}$ is strictly dominated by $a_k$ for player 1. Therefore, every action in $X' \setminus \{a_1, a_2, \ldots, a_m\}$ will not be chosen with positive probability in an equilibrium for player 1. As a result, only the actions in $X'' = \{a_1, a_2, \ldots, a_m\}$ can be chosen with positive probabilities in an equilibrium for player 1.

Similarly, only the actions in $X'' = \{a_1, a_2, \ldots, a_m\}$ can be chosen with positive probabilities in an equilibrium for player 2.

\bigskip

We can focus on the case that the two players only choose actions from the common action space $X'' = \{a_1, a_2, \ldots, a_m\}$. The payoff matrix restricted on the action set $X''$ is as follows.
\begin{figure}[ht]\hspace*{\fill}%
\begin{game}{5}{5}[Player $1$][Player $2$][]
		       & $a_1$	    & $a_2$         & $a_3$	        & $\cdots$		& $a_m$		\\
$a_1$	       & $1,-1$	    & $-1,1$	    & $0,0$	        & $\cdots$      & $0,0$	\\
$a_2$	       & $0,0$	    & $1,-1$	    & $-1,1$	    & $\cdots$      & $0,0$	\\
$a_3$	       & $0,0$	    & $0,0$	        & $1,-1$	    & $\cdots$      & $0,0$	\\
$\vdots$	   & $\vdots$	& $\vdots$	    & $\vdots$	    & $\vdots$	    & $\vdots$	\\
$a_m$	       & $-1,1$	    & $0,0$         & $\cdots$	    & $0,0$	        & $1,-1$	\\
\end{game}\hspace*{\fill}%
\end{figure}
\FloatBarrier
\noindent In each cell, the first number is the payoff for player 1 and the second number is the payoff for player 2.

\paragraph{Step 2}

Next, we construct a new auxiliary $2$-player game $\Gamma_2$. Recall that $\tau$ is the uniform distribution on the triangle $\{ (l_1, l_2) \mid 0 \le l_1 \le l_2 \le 1 \}$. We use $\tau_i$ to denote the marginal of $\tau$ on $L_i$. Let
$$ q(t_1, t_2) = \begin{cases}
\frac{1}{2(1-\phi_1(t_1))\phi_2(t_2)},  & \text{if } 0 < \phi_1(t_1) \le \phi_2(t_2) < 1,   \\
0,                                      & \text{otherwise},
\end{cases} $$
where $\phi_i$ is a measure-preserving mapping from $(T_i, \cF_i, \lambda_i)$ to $([0, 1], \cB, \tau_i)$ such that for any $E \in \cF_i$ there exists a set $E' \in \cB$ with $\lambda_i \big( E \Delta \phi_i^{-1}(E') \big) = 0$.

The components of game $\Gamma_2$ is as follows: (1) Players~$1$ and $2$'s action spaces and payoffs are the same as in the game $G_2$; (2) The private type space for each player $i$ is $(T_i, \cT_i, \lambda_i)$; (3) The common prior $\lambda$ has the Radon-Nikodym derivative $q$ with respect to $\lambda_1 \otimes \lambda_2$.

It can be easily checked that each $\lambda_i$ is the marginal of $\lambda$ on $T_i$. Based on the analysis in Step 1, both players only choose actions from the set $X''$ in an equilibrium. Then $\Gamma_2$ is reduced to the game constructed in \citet[p.31]{HS2019}. Following the arguments therein, we have that $\cT_i$ is nowhere equivalent to $\cF_i$ under $\lambda_i$ for each $i = 1, 2$. Thus, players 1 and 2 have decomposable coarser payoff-relevant information.

The proof is then completed by adding dummy players. In particular, one can consider an $n$-player game in which only players~$1$ and $i$ are active for some $2 \le i \le n$, while all other players are inactive. The payoffs, action sets and private type spaces of players $1$ and $i$ are the same as those of players $1$ and $2$ in $\Gamma_2$. Then the above argument shows that players $1$ and $i$ have decomposable coarser payoff-relevant information. This further implies that all the players have decomposable coarser payoff-relevant information.



\subsection{Proof of Theorem~\ref{thm:purification}}
\label{subsec:proof:purification-suff}

In the following two subsections, we provide the proofs for the sufficiency part and necessity part of Theorem~\ref{thm:purification}, respectively.

\subsubsection{Proof for the sufficiency part of Theorem~\ref{thm:purification}}

To prove the sufficiency part of Theorem~\ref{thm:purification}, we first present a new purification result under a vector measure.

\begin{lem}
\label{lem:conditional-purify}
Let $\mu_0, \mu_1, \ldots, \mu_J$ be probability measures on some measurable space $(T, \cT)$ such that $\mu_j$ is absolutely continuous with respect to $\mu_0$ with bounded density for $j = 1, 2, \ldots, J$. Let $A$ be a compact metric space, and $\{v_k\}_{k=1}^K$ be $\cB(A) \otimes \cF$-measurable mappings from $A \times T$ to $\bR$ such that $v_k$ is integrably bounded under $\mu_0$ for each $a \in A$ and $k = 1, 2, \ldots, K$. If $\cT$ is nowhere equivalent to $\cF$ under $\mu_0$, then for any $\cT$-measurable transition probability $g$ from $T$ to $\cM(A)$, there exists a $\cT$-measurable mapping $f$ from $T$ to $A$ such that for $j = 0, 1, \ldots, J$,
\begin{enumerate}
\item for $k =1, 2, \ldots, K$,
    $$ \int_T \int_A  v_k(a, t)  g(t; \dif a)  \mu_j(\dif t)  =  \int_T  v_k(f(t), t)  \mu_j(\dif t); $$

\item for any $D \in \cF$ and any $B \in \cB(A)$,
    $$ \int_D  g(t; B)  \mu_j(\dif t)  =  \int_D  \mathbf{1}_B \bigl(f(t)\bigr)  \mu_j(\dif t); $$

\item for any $\cF$-measurable mapping $g_1$ from $T$ to some Polish space $Y$ and $k = 1, 2, \ldots, K$,
    $$ \mu_j \circ (g, g_1)^{-1}  =  \mu_j \circ (f, g_1)^{-1}, $$
    where $\mu_j \circ (g, g_1)^{-1}$ and $\mu_j \circ (f, g_1)^{-1}$ denote the joint distributions on $A \times Y$:
    $$ \mu_j \circ (g, g_1)^{-1}(B \times B_1)  =  \int_{\{t \in T \mid g_1(t) \in B_1\}}  g(t; B)  \mu_j(\dif t) $$
    and
    $$ \mu_j \circ (f, g_1)^{-1}(B \times B_1)  =  \int_{\{t \in T \mid g_1(t) \in B_1\}}  \mathbf{1}_{B}\bigl(f(t)\bigr)  \mu_j(\dif t) $$
    for any $B \in \cB(A)$ and $B_1 \in \cB(Y)$.

\item $f(t) \in \supp g(t)$ for $\mu_0$-almost all $t \in T$.
\end{enumerate}
\end{lem}

\begin{proof}
By Lemma~\ref{lem:rcd}, there exists a mapping $f$ from $T$ to $A$ such that
$$ \int_D  g(t; B)  \mu_j(\dif t)  =  \int_D  \mathbf{1}_{B} \big(f(t)\big)  \mu_j(\dif t) $$
for any $D \in \cF$, $B \in \cB(A)$, and $j = 0, 1, \ldots, J$. This proves part 2.

This equality further implies that
$$ \int_T  \int_A  \mathbf{1}_D(t)  \mathbf{1}_B(a)  g(t; \dif a)  \mu_j(\dif t)  =  \int_T  \mathbf{1}_D(t)  \mathbf{1}_B\bigl(f(t)\bigr)  \mu_j(\dif t). $$
Fix a $ \cB(A) \otimes \cF$-measurable mapping $v_k$. Without loss of generality, we assume that it is nonnegative. Then $v_k$ is an increasing limit of a sequence of simple functions. By the monotone convergence theorem, we have that
$$ \int_T  \int_A  v_k(a, t)  g(t; \dif a)  \mu_j(\dif t)  =  \int_T  v_k\bigl(f(t), t\bigr)  \mu_j(\dif t). $$
This proves part 1.

In addition, part 4 is shown via replacing $v_k$ by the mapping $c(a, t) = \mathbf{1}_{\supp g(t)}(a)$. Part 3 follows by noting that the set $\{t \in T \mid g_1(t) \in B_1\} \in \cF$ for any $\cF$-measurable mapping $g_1$ from $T$ to some Polish space $Y$ and $B_1 \in \cB(Y)$.
\end{proof}

\begin{rmk}
\rm
The above lemma covers several purification results in the literature. \cite{Sun1996} and \cite{LS2006}\footnote{The proof of \citet[Theorem~2.2]{LS2006} needs to be slightly modified as follows. Let $\cD = \{\phi_n\}_{n \ge 1}$. In the third line of p.~751, let $\nu$ be the vector measure on $(T, \cT)$ with values in $\bR^{m k_m}$ for which the $(i-1) k_m + k$-th component is $\nu_k^{\phi_i, m}$. Note that for each $\phi \in \cD$, there exists a positive integer $M_{\phi}$ such that $\phi = \phi_{M_{\phi}}$.
The rest of the argument would go through by working with $m \ge M_{\phi}$ on p.~751.} worked with atomless Loeb spaces as the probability spaces, which was extended to saturated probability spaces in \cite{Podczeck2009} and \cite{WZ2012}. \cite{HS2014} presented a purification result based on a probability measure, which generalizes these earlier results. Lemma~\ref{lem:conditional-purify} above obtains a further generalization in the setting with vector measures.
\end{rmk}

\begin{proof}[Proof for the sufficiency part of Theorem~\ref{thm:purification}]

Recall that the measure $\mu_{ij}$ on $(T_i, \cT_i)$ is absolutely continuous with respect to $\lambda_i$ with the density $\rho_i^j$ for $j = 1, 2, \ldots, J$, $\mu_{i0} = \lambda_i$, and $\mu_i = (\mu_{i0}, \mu_{i1}, \ldots, \mu_{iJ})$. Fix a behavioral-strategy profile $g$. For each $i \in I$, $k = 1, 2, \ldots, J$, $a_i \in A_i$, and $t_i \in T_i$, let
$$ V_{ik}^g (a_i, t_i)  =  \int_{T_{-i}}  \int_{A_{-i}}  w^k_i(a_i, a_{-i}, t_i, t_{-i})  \prod_{\ell \ne i}  g_\ell(t_\ell; \dif a_\ell)  \mu_{-ik} (\dif t_{-i}). $$

By Lemma~\ref{lem:conditional-purify}, for each $i \in I$, there exists a $\cT_i$-measurable mapping $f_i \colon T_i \to A_i$ such that for each $j = 0, 1, \ldots, J$,
\begin{enumerate}[label=(\arabic*)]
\item for each $k = 1, 2, \ldots, J$,
    $$ \int_{T_i}  \int_{A_i}  V_{ik}^g(a_i, t_i)  g_i(t_i; \dif a_i)  \mu_{ij}(\dif t_i)  =  \int_{T_i}  V_{ik}^g(f_i(t_i), t_i)  \mu_{ij}(\dif t_i); $$

\item for each $D \in \cF_i$ and $B \in \cB(A_i)$,
    $$ \int_D  g_i(t_i; B)  \mu_{ij} (\dif t_i)  =  \int_D  \mathbf{1}_{B}(f_i(t_i))  \mu_{ij} (\dif t_i); $$

\item $f_i(t_i) \in \supp g_i(t_i)$ for $\lambda_i$-almost all $t_i \in T_i$.
\end{enumerate}
Then (2) implies that $f$ and $g$ are conditional distribution equivalent, and (3) means that $f$ and $g$ are belief consistent.

Given $i \in I$, $j = 1, 2, \ldots, J$, $t_i \in T_i$ and $a_i \in A_i$, we have that
\begin{align}
\label{eq:V}
V_{ij}^g(a_i, t_i)  &  =  \int_{T_{-i}}  \int_{A_{-i}}  w^j_i(a_i, a_{-i}, t_i, t_{-i})  \prod_{\ell \ne i} g_\ell(t_\ell; \dif a_\ell)  \mu_{-ij} (\dif t_{-i})    \nonumber\\
                    &  =  \int_{T_{-i}}  \int_{A_{-i}}  w^j_i(a_i, a_{-i}, t_i, t_{-i})  \prod_{\ell \ne i} \mu_{\ell j}^{f_\ell | \cF_\ell}(t_\ell; \dif a_\ell)  \mu_{-ij} (\dif t_{-i})  \nonumber\\
                    &  =  \int_{T_{-i}}  w^j_i(a_i, f_{-i}(t_{-i}), t_i, t_{-i})  \mu_{-ij} (\dif t_{-i})   \nonumber\\
                    &  =  V_{ij}^f(a_i, t_i).
\end{align}
The second equality holds because of Property~(2) above, and the third equality is due to Equation~\eqref{eq:epayoff}. Thus, for any behavioral strategy $h_i$ and $t_i \in T_i$,
\begin{equation}
\label{eq:V2}
\int_{A_i}  V_{ij}^f(a_i, t_i)  h_i(t_i; \dif a_i)  =  \int_{A_i}  V_{ij}^g(a_i, t_i)  h_i(t_i; \dif a_i).
\end{equation}
We have that
\begin{align*}
U_i(g)  &  =  \sum_{j=1}^J  \int_{T_i}  \int_{A_i}  \rho^j_i (t_i)  V_{ij}^g (a_i, t_i)  g_i(t_i; \dif a_i)  \lambda_i (\dif t_i)   \\
        &  =  \sum_{j=1}^J  \int_{T_i}  \int_{A_i}  V_{ij}^g (a_i, t_i)  g_i(t_i; \dif a_i)  \mu_{ij} (\dif t_i)    \\
        &  =  \sum_{j=1}^J  \int_{T_i}  V_{ij}^g (f_i(t_i), t_i)  \mu_{ij} (\dif t_i)   && \text{due to Property~(1)}   \\
        &  =  \sum_{j=1}^J  \int_{T_i}  V_{ij}^f (f_i(t_i),t_i)  \mu_{ij} (\dif t_i)  =  U_i(f),    && \text{due to Equation~\eqref{eq:V}}
\end{align*}
and
\begin{align*}
U_i(h_i, g_{-i})    &  =  \sum_{j=1}^J  \int_{T_i}  \int_{A_i}  \rho^j_i (t_i)  V_{ij}^g (a_i, t_i)  h_i(t_i; \dif a_i)  \lambda_i (\dif t_i)   \\
                    &  =  \sum_{j=1}^J  \int_{T_i}  \int_{A_i}  V_{ij}^g (a_i, t_i)  h_i(t_i; \dif a_i)  \mu_{ij} (\dif t_i)    \\
                    &  =  \sum_{j=1}^J  \int_{T_i}  \int_{A_i}  V_{ij}^f (a_i, t_i)  h_i(t_i; \dif a_i)  \mu_{ij} (\dif t_i)  =  U_i(h_i, f_{-i}).  && \text{due to Equation~\eqref{eq:V2}}
\end{align*}
Thus, $f$ and $g$ are payoff equivalent. Therefore, $f$ is a conditional purification of $g$.
\end{proof}

\subsubsection{Proof for the necessity part of Theorem~\ref{thm:purification}}

Below, we present an equivalence result for the notion of nowhere equivalence (see Lemma~2 in \cite{HS2021}), which is useful for deriving the necessity part of Theorem~\ref{thm:purification}.

\begin{lem}
\label{lem:noeq}
The following two conditions are equivalent.
\begin{itemize}
\item The $\sigma$-algebra $\cT$ is nowhere equivalent to $\cF$ under a probability measure $\mu$.
\item The sub-$\sigma$-algebra $\cF$ admits an asymptotic independent supplement in $\cT$ under $\mu$; that is, for some strictly increasing sequence $\{k_m\}_{m=1}^\infty$ and each $m \ge 1$, there exists a $\cT$-measurable partition $\{E_1, E_2, \ldots, E_{k_m}\}$ of $T$ such that for $j = 1, 2, \ldots, k_m$, (a) $\mu(E_j) = \frac{1}{k_m}$, and (b)~$E_j$ is independent of $\cF$ under $\mu$.
\end{itemize}
\end{lem}

\begin{proof}[Proof for the necessity part of Theorem~\ref{thm:purification}]
Fix a positive integer $m \ge 2$. We consider the following game. The set of players is $I = \{ 1, 2, \ldots, n \}$. Each player $i$ has the private type space $(T_i, \cT_i, \lambda_i)$. The common prior is $\lambda = \otimes_{i \in I} \lambda_i$. The common action space is $A = [0, m]$.

The payoffs are defined as follows. Let $([0,1], \cB, \eta)$ be the Lebesgue unit interval, where $\cB$ is the Borel $\sigma$-algebra on $[0, 1]$ and $\eta$ is the Lebesgue measure. As is well known, for each $i \in I$, there is a measure-preserving mapping $\phi_i$ from $(T_i, \cF_i, \lambda_i)$ to $([0, 1], \cB, \eta)$ such that for any $E \in \cF_i$, there exists a set $E' \in \cB$ with $\lambda_i \bigl (E \triangle \phi_i^{-1}(E') \bigr) = 0$. For the action profile $(a_1, a_2, \ldots, a_n) \in \prod_{i \in I} A$ and type profile $(t_1, t_2, \ldots, t_n) \in \prod_{i \in I} T_i$, the payoff of player~$i$ is given by
$$ u_i (a_1, a_2, \ldots, a_n, t_1, t_2, \ldots, t_n)  =  - \prod_{j=0}^{m-1}  \big[ a_i - \phi_i(t_i) - j \big]^2.$$
Notice that the payoff of player~$i$ does not depend on the actions and the types of her opponents.

Define a behavioral strategy $g_1 \colon T_1 \to \cM(A)$ for player~$1$ as
$$ g_1(t_1)  =  \frac{1}{m}  \sum_{j=0}^{m-1}  \delta_{\phi_1(t_1) + j}, $$
where $\delta_{\phi_1(t_1) + j}$ is the Dirac measure on $A$ at the points $\phi_1(t_1) + j$. Let $\hat{g}_1 \colon [0, 1] \to \cM(A)$ be a function on the unit interval as
$$ \hat{g}_1(l_1)  =  \frac{1}{m}  \sum_{j=0}^{m-1}  \delta_{l_1 + j}. $$
Then $g_1 = \hat{g}_1 \circ \phi_1$. Let $\tau_1$ be a probability measure on $A$ such that for any Borel subset $B$ in $A$,
$$ \tau_1(B)  =  \int_{T_1}  g_1(t_1; B)  \lambda_1(\dif t_1), $$
which implies that
$$ \tau_1(B)  =  \int_{T_1}  \hat{g}_1 (\phi_1(t_1); B)  \lambda_1(\dif t_1)  =  \int_{[0,1]} \hat{g}_1 (l_1; B)  \eta(\dif l_1). $$
Given the definition of $\hat{g}_1$, it is clear that $\tau$ is the uniform distribution on $A$.

Due to the assumption of necessity part of Theorem~\ref{thm:purification}, the behavioral strategy $g_1$ has a conditional purification $f_1$. By the definition, $f_1$ and $g_1$ are distribution equivalent. As a result, $\lambda_1 f_1^{-1}(B) = \int_{T_1} g_1(t_1; B) \lambda_1(\dif t_1) = \tau(B)$; that is, $\lambda_1 f_1^{-1} = \tau$. Based on the belief consistency condition, $f_1(t_1) \in \{\phi_1(t_1), \phi_1(t_1) + 1, \ldots, \phi_1(t_1) + m-1\}$ for $\lambda_1$-almost all $t_1 \in T_1$. Thus, $f_1(t_1) = \phi_1(t_1) + j$ on a $\cT_1$-measurable set $C_j \subseteq T_1$ for $j = 0, 1, \ldots, m-1$.

Recall that for any $E \in \cF_1$, there exists a set $E' \in \cB$ with $\lambda_1 \bigl( E \triangle \phi_1^{-1}(E') \bigr) = 0$. Then for each $j = 0, 1, \ldots, m-1$, we have that
\begin{align*}
\lambda_1(C_j \cap E)   &  =  \lambda_1 \bigl( C_j \cap \phi_1^{-1}(E') \bigr)  =  \lambda_1(f_1 \in E' + j)    \\
                        &  =  \tau_1(E' + j)  =  \frac{1}{m} \eta(E')  =  \frac{1}{m} \lambda_1 \bigl( \phi_1^{-1}(E') \bigr)  =  \frac{1}{m} \lambda_1(E),
\end{align*}
where $E' + j$ denotes the set $\{l + j \mid l \in E'\}$. Thus, we have that $\lambda_1(C_j) = \frac{1}{m}$. Therefore, $\{C_0, C_1, \ldots, C_{m-1}\}$ is a $\cT_1$-measurable partition of $T_1$ such that $C_j$ is independent of $\cF_1$ for $j = 0, 1, \ldots, m-1$. Since $m$ is arbitrary, $\cF_1$ admits an asymptotic independent supplement in $\cT_1$ under $\lambda_1$. By Lemma~\ref{lem:noeq}, $\cT_1$ is nowhere equivalent to $\cF_1$ under $\lambda_1$. Based on an analogous argument, $\cT_i$ is nowhere equivalent to $\cF_i$ under $\lambda_i$ for each $i \in I$.
\end{proof}

\subsection{Proofs of Claims~\ref{claim-example 1} and \ref{claim:allpay}}
\label{subsec:proof:claim}

\begin{proof}[Proof of Claim~\ref{claim-example 1}]

We first explicitly give the marginal $\lambda_i$ of $\lambda$, $i = 1, 2$. For any $D \in \cB([0,1])$, we have
\begin{align*}
\lambda_1(D)
& = \frac{1}{2}\int_{D}\int_{T_2} 1 \; \eta(\rmd t_2) \eta(\rmd t_1) + \frac{1}{2}\int_{D}\int_{T_2} 6t_1t^2_2 \eta(\rmd t_2) \eta(\rmd t_1) \\
& = \int_{D} (\frac{1}{2} + t_1) \eta(\rmd t_1).
\end{align*}
The density of $\lambda_1$ with respect to $\eta$ is $\frac{1}{2} + t_1$. Similarly, the density of $\lambda_2$ on $\eta$ is $\frac{1}{2} + \frac{3}{2}t^2_2$.

We need to verify that $\lambda$ is absolutely continuous with respect to $\lambda_1 \otimes \lambda_2$. For any $D_1, D_2 \in \cB([0,1])$,
\begin{align*}
\lambda(D_1 \times D_2)
& = \frac{1}{2}\int_{D_1}\int_{D_2} 1 \; \eta(\rmd t_2) \eta(\rmd t_1) + \frac{1}{2}\int_{D_1}\int_{D_2} 6t_1t^2_2 \eta(\rmd t_2) \eta(\rmd t_1) \\
& = \int_{D_1}\int_{D_2} \frac{\frac{1}{2} + 3t_1t^2_2}{(\frac{1}{2} + t_1)(\frac{1}{2} + \frac{3}{2}t^2_2)} (\frac{1}{2} + t_1)(\frac{1}{2} + \frac{3}{2}t^2_2) \eta(\rmd t_2) \eta(\rmd t_1) \\
& = \int_{D_1}\int_{D_2} \frac{\frac{1}{2} + 3t_1t^2_2}{(\frac{1}{2} + t_1)(\frac{1}{2} + \frac{3}{2}t^2_2)} \lambda_2(\rmd t_2) \lambda_1(\rmd t_1).
\end{align*}
That is, $\lambda$ has the density $q(t_1, t_2) =  \frac{\frac{1}{2} + 3t_1t^2_2}{(\frac{1}{2} + t_1)(\frac{1}{2} + \frac{3}{2}t^2_2)}$ on $\lambda_1 \otimes \lambda_2$. The mapping $q(t_1, \cdot)$ induces the full $\sigma$-algebra $\cB([0,1])$ on $T_2$ when $t_1 > \frac{1}{2}$, as $q(t_1, \cdot)$ is continuous and strictly increasing in $t_2$. Recall that $u_1$ only depends on the action profile $a$. Thus, $w_1(a, t_1, \cdot)$ also induces $\cB([0,1])$ on $T_2$ when $t_1 > \frac{1}{2}$, implying that DCPI is not satisfied for $J = 1$.

One can write
$$w_1(a, t_1, t_2) = \frac{1}{2}u_1(a) \frac{1}{\frac{1}{2} + t_1}\frac{1}{\frac{1}{2} + \frac{3}{2}t^2_2} + 3u_1(a) \frac{t_1}{\frac{1}{2} + t_1}\frac{t^2_2}{\frac{1}{2} + \frac{3}{2}t^2_2}.
$$
Let $w_1^1(a,t) = \frac{1}{2}u_1(a)$, $\rho_1^1(t_1) = \frac{1}{\frac{1}{2} + t_1}$, $\rho_2^1(t_2) =\frac{1}{\frac{1}{2} + \frac{3}{2}t^2_2}$, $w_1^2(a, t) = 3u_1(a)$, $\rho_1^2(t_1) = \frac{t_1}{\frac{1}{2} + t_1}$, $\rho_2^2(t_2) = \frac{t^2_2}{\frac{1}{2} + \frac{3}{2}t^2_2}$. Note that $w_1^1$ and $w_1^2$ do not depend on the type profile $t$. By Lemma~3 in \cite{HS2019}, there exists a sub-$\sigma$-algebra $\cG_i \subseteq \cT_i$ such that $(T_i, \cG_i, \lambda_i)$ is atomless and $\cT_i$ is nowhere equivalent to $\cG_i$ for $i = 1, 2$. Thus,  DCPI is satisfied for $J = 2$. For the case $J > 2$, one simply lets $\rho_i^j \equiv 0$ for $j > 2$ and $i = 1, 2$. This completes the proof.
\end{proof}

\medskip

\begin{proof}[Proof of Claim~\ref{claim:allpay}]

We first check that the condition of uniform payoff security is satisfied.

Recall that $\psi_1$, $\psi_2$, and $\{\varphi_i\}_{i \in I}$ are all continuous on the compact set $\tilde{T} \times A$. For any $\epsilon > 0$, there exists some $\delta > 0$ such that for any $(\tilde{t}, a), (\tilde{t}', a') \in \tilde{T} \times A$ with $\|(\tilde{t}, a) - (\tilde{t}', a')\|_2 < \delta$, and for each $i \in I$,
$$ | \psi_1(\tilde{t}, a) - \psi_1(\tilde{t}', a') |  <  \tfrac{\epsilon}{3}, \quad | \psi_2(\tilde{t}, a) - \psi_2(\tilde{t}', a') |  <  \tfrac{\epsilon}{3}, \quad | \varphi_i(\tilde{t}, a) - \varphi_i(\tilde{t}', a') | < \tfrac{\epsilon}{3}, $$
where $\|\cdot\|_2$ denotes the Euclidean norm in $\bR^3$.

Fix a bidding strategy profile $(\bm{b}_1, \bm{b}_2, \ldots, \bm{b}_n)$. For each bidder~$i$, we construct a new bidding strategy as follows: $\bm{b}_i^*(t_i) = \min\{\bm{b}_i(t_i) + \frac{\delta}{2}, \bar{a}\}$. To verify the condition of uniform payoff security, we shall show that for each $i \in I$ and for all $(t, a_{-i})$, there exists a neighborhood $O_{a_{-i}}$ of $a_{-i}$ such that for all $y_{-i} \in O_{a_{-i}}$,
$$ u_i(t, \bm{b}_i^*(t_i), y_{-i}) - u_i(t, \bm{b}_i(t_i), a_{-i}) > - \epsilon. $$

Suppose that bidder~$i$ follows the bidding strategy $\bm{b}_i$. Fix the type profile $t$ and other bidders' bids $a_{-i}$. There are three cases to consider.

\paragraph{Case 1}

Bidder~$i$ is a losing bidder. In this case, $\bm{b}_i(t_i) < \max_{k \neq i} a_k$. Bidder~$i$' payoff is
$$ u_i(t, \bm{b}_i(t_i), a_{-i})  =  \sum_{j=1}^J \Bigl( \psi_2(t^{j}_{0}, t_{11}, \ldots, t_{n1}, \bm{b}_i(t_i), a_{-i}) + \varphi_i(t^{j}_{0}, t_{11}, \ldots, t_{n1}, \bm{b}_i(t_i), a_{-i}) \Bigr)  \kappa(\{t^j_0\} \mid t). $$

Let $O_{\frac{\delta}{2}}(a_{-i})$ be the ball in $A_{-i}$ with center $a_{-i}$ and radius $\frac{\delta}{2}$. For any $y_{-i} \in O_{\frac{\delta}{2}}(a_{-i})$, $\|(\bm{b}_i(t_i), a_{-i}) - (\bm{b}_i^*(t_i), y_{-i})\|_2 < \delta$. Thus, for each $j = 1, 2, \ldots, J$,
\begin{align*}
\psi_1(t^j_0, t_{11}, \ldots, t_{n1}, \bm{b}_i^*(t_i), y_{-i})  & \ge \psi_2(t^j_0, t_{11}, \ldots, t_{n1}, \bm{b}_i^*(t_i), y_{-i})    \\
                                                                & > \psi_2(t^j_0, t_{11}, \ldots, t_{n1}, \bm{b}_i(t_i), a_{-i}) - \tfrac{\epsilon}{3},
\end{align*}
and
$$ \varphi_i(t^j_0, t_{11}, \ldots, t_{n1}, \bm{b}_i^*(t_i), y_{-i})  \ge  \varphi_i(t^j_0, t_{11}, \ldots, t_{n1}, \bm{b}_i(t_i), a_{-i}) - \tfrac{\epsilon}{3}. $$
For any $y_{-i} \in O_{\frac{\delta}{2}}(a_{-i})$, the two inequalities above imply that
\begin{align*}
        & u_i(t, \bm{b}_i^*(t_i), y_{-i})   \\
\ge{}   & \sum_{j=1}^J  \Big( \psi_2(t^j_0, t_{11}, \ldots, t_{n1}, \bm{b}_i^*(t_i), y_{-i}) + \varphi_i(t^j_0, t_{11}, \ldots, t_{n1}, \bm{b}_i^*(t_i), y_{-i}) \Big)  \kappa(\{t^j_0\} \mid t)    \\
>{}     & \sum_{j=1}^J  \Big( \psi_2(t^j_0, t_{11}, \ldots, t_{n1}, \bm{b}_i(t_i), a_{-i}) + \varphi_i(t^j_0, t_{11}, \ldots, t_{n1}, \bm{b}_i(t_i), a_{-i}) \Big)  \kappa(\{t^j_0\} \mid t)  -  \tfrac{2\epsilon}{3}   \\
>       & u_i(t, \bm{b}_i(t_i), a_{-i}) - \epsilon.
\end{align*}

\paragraph{Case 2}

Bidder~$i$ is the unique winning bidder. In this case, $\bm{b}_i(t_i) > \max_{k \neq i} a_k$. Bidder~$i$' payoff is
$$ u_i(t, \bm{b}_i(t_i), a_{-i})  =  \sum_{j=1}^J  \Big( \psi_1(t^j_0, t_{11}, \ldots, t_{n1}, \bm{b}_i(t_i), a_{-i}) + \varphi_i(t^j_0, t_{11}, \ldots, t_{n1}, \bm{b}_i(t_i), a_{-i}) \Big)  \kappa(\{t^j_0\} \mid t). $$

Let $O_{\delta'}(a_{-i})$ be the $\delta'$-ball of $a_{-i}$ such that $\delta' < \frac{\delta}{2}$ and
$$ \bm{b}_i(t_i)  >  \max_{y_{-i} \in O_{\delta'}(a_{-i}), k \ne i} y_k. $$
Since $\bm{b}_i^*(t_i) \ge \bm{b}_i(t_i)$, bidder~$i$ is still the unique winner by adopting the bidding strategy $\bm{b}_i^*$ if others bid $y_{-i} \in O_{\delta'}(a_{-i})$. For any $y_{-i} \in O_{\delta'}(a_{-i})$, since $\|(\bm{b}_i(t_i), a_{-i}) - (\bm{b}_i^*(t_i), y_{-i})\|_2 < \delta$,
$$ \psi_1(t^j_0, t_{11}, \ldots, t_{n1}, \bm{b}_i^*(t_i), y_{-i})  >  \psi_1(t^j_0, t_{11}, \ldots, t_{n1}, \bm{b}_i(t_i), a_{-i})  -  \tfrac{\epsilon}{3} $$
for each $j = 1, 2, \ldots, J$. It implies that
$$ u_i(t, \bm{b}_i^*(t_i), y_{-i}) - u_i(t, \bm{b}_i(t_i), a_{-i})  >  - \epsilon. $$

\paragraph{Case 3}

Bidder~$i$ is one of the winning bidders. In this case, $\bm{b}_i(t_i) = \max_{\ell \ne i} a_\ell$. Bidder~$i$' payoff is
\begin{align*}
    & u_i(t, \bm{b}_i(t_i), a_{-i}) \\
={} & \sum_{j=1}^J  \bigg[ \frac{\xi_i(\bm{b}_i(t_i), a_{-i})}{\xi_i(\bm{b}_i(t_i), a_{-i}) + \sum_{\ell \ne i \colon a_\ell = \bm{b}_i(t_i)} \xi_\ell(\bm{b}_i(t_i), a_{-i})} \psi_1(t^{j}_{0}, t_{11}, \ldots, t_{n1}, \bm{b}_i(t_i), a_{-i})  \\
    & + \Big( 1 - \frac{\xi_i(\bm{b}_i(t_i), a_{-i})}{\xi_i(\bm{b}_i(t_i), a_{-i}) + \sum_{\ell \ne i \colon a_\ell = \bm{b}_i(t_i)} \xi_\ell(\bm{b}_i(t_i), a_{-i})} \Big) \psi_2(t^{j}_{0}, t_{11}, \ldots, t_{n1}, \bm{b}_i(t_i), a_{-i})    \\
    & + \varphi_i(t^{j}_{0}, t_{11}, \ldots, t_{n1}, \bm{b}_i(t_i), a_{-i}) \bigg] \kappa(\{t^j_0\} \mid t).
\end{align*}

If bidder~$i$ becomes the unique winner by bidding $\bm{b}_i^*(t_i)$ when others bid $a_{-i}$, then $\bm{b}_i^*(t_i) > \bm{b}_i(t_i)$. One can identify a neighbourhood $O_{\delta'}(a_{-i})$ of $a_{-i}$ such that $\delta' < \frac{\delta}{2}$ and
$$ \bm{b}_i^*(t_i)  >  \max_{y_{-i} \in O_{\delta'}(a_{-i}), \ell \ne i} y_\ell. $$
For any $y_{-i} \in O_{\delta'}(a_{-i})$, since $\|(\bm{b}_i(t_i), a_{-i}) - (\bm{b}_i^*(t_i), y_{-i})\|_2 < \delta$,
\begin{align*}
        & u_i(t, \bm{b}_i^*(t_i), y_{-i})   \\
={}     & \sum_{j=1}^J  \bigg[ \psi_1(t^j_0, t_{11}, \ldots, t_{n1}, \bm{b}_i^*(t_i), y_{-i}) + \varphi_i(t^j_0, t_{11}, \ldots, t_{n1}, \bm{b}_i^*(t_i), y_{-i}) \bigg]  \kappa(\{t^j_0\} \mid t)    \\
={}     & \sum_{j=1}^J  \bigg[ \frac{\xi_i(\bm{b}_i(t_i), a_{-i})}{\xi_i(\bm{b}_i(t_i), a_{-i}) + \sum_{\ell \ne i \colon a_\ell = \bm{b}_i(t_i)} \xi_\ell(\bm{b}_i(t_i), a_{-i})} \psi_1(t^j_0, t_{11}, \ldots, t_{n1}, \bm{b}_i^*(t_i), y_{-i})  \\
        & + \Big( 1 - \frac{\xi_i(\bm{b}_i(t_i), a_{-i})}{\xi_i(\bm{b}_i(t_i), a_{-i}) + \sum_{\ell \ne i \colon a_\ell = \bm{b}_i(t_i)} \xi_\ell(\bm{b}_i(t_i), a_{-i})} \Big) \psi_1(t^j_0, t_{11}, \ldots, t_{n1}, \bm{b}_i^*(t_i), y_{-i})    \\
        & + \varphi_i(t^j_0, t_{11}, \ldots, t_{n1}, \bm{b}_i^*(t_i), y_{-i}) \bigg]  \kappa(\{t^j_0\} \mid t)   \\
>{}     & \sum_{j=1}^J  \bigg[ \frac{\xi_i(\bm{b}_i(t_i), a_{-i})}{\xi_i(\bm{b}_i(t_i), a_{-i}) + \sum_{\ell \ne i \colon a_\ell = \bm{b}_i(t_i)} \xi_\ell(\bm{b}_i(t_i), a_{-i})} \psi_1(t^j_0, t_{11}, \ldots, t_{n1}, \bm{b}_i(t_i), a_{-i})    \\
        & + \Big( 1 - \frac{\xi_i(\bm{b}_i(t_i), a_{-i})}{\xi_i(\bm{b}_i(t_i), a_{-i}) + \sum_{\ell \ne i \colon a_\ell = \bm{b}_i(t_i)} \xi_\ell(\bm{b}_i(t_i), a_{-i})} \Big) \psi_2(t^j_0, t_{11}, \ldots, t_{n1}, \bm{b}_i(t_i), a_{-i})  \\
        & + \varphi_i(t^j_0, t_{11}, \ldots, t_{n1}, \bm{b}_i(t_i), a_{-i}) \bigg]  \kappa(\{t^j_0\} \mid t) - \tfrac{2\epsilon}{3} \\
>{}     & u_i(t, \bm{b}_i(t_i), a_{-i}) - \epsilon.
\end{align*}

If bidder~$i$ is still one of winning bidders by bidding $\bm{b}_i^*(t_i)$ when others bid $a_{-i}$, then $\bm{b}_i^*(t_i) = \bm{b}_i(t_i) = \bar{a}$. That is, no matter what the other bidders bid, bidder~$i$ must be a winning bidder by bidding $\bm{b}_i^*(t_i)$. Pick $\delta'' < \delta'$ such that for any $y_{-i} \in O_{\delta''}(a_{-i})$, (1) $y_\ell < \bar{a}$ if $a_\ell < \bar{a}$ for $\ell \ne i$, and (2)
$$ \bigg| \frac{\xi_i(\bm{b}_i^*(t_i), y_{-i})}{\xi_i(\bm{b}_i^*(t_i), y_{-i}) + \sum\limits_{\ell \ne i: a_\ell = \bar{a}} \xi_\ell(\bm{b}_i^*(t_i), y_{-i})} - \frac{\xi_i(\bm{b}_i(t_i), a_{-i})}{\xi_i(\bm{b}_i(t_i), a_{-i}) + \sum\limits_{\ell \ne i: a_\ell = \bar{a}} \xi_\ell(\bm{b}_i(t_i), a_{-i})} \bigg| < \epsilon_1, $$
where $0 < \epsilon_1 < \frac{\xi_i(\bm{b}_i(t_i), a_{-i})}{\xi_i(\bm{b}_i(t_i), a_{-i}) + \sum\limits_{\ell \ne i: a_\ell = \bar{a}} \xi_\ell(\bm{b}_i(t_i), a_{-i})}$ and
$$ \epsilon_1 \sum_{j=1}^J \Big( \psi_1(t^j_0, t_{11}, \ldots, t_{n1}, \bm{b}_i(t_i), a_{-i}) - \psi_2(t^{j}_{0}, t_{11}, \ldots, t_{n1}, \bm{b}_i(t_i), a_{-i}) \Big) \kappa(\{t^j_0\} \mid t) < \tfrac{\epsilon}{3}. $$
For any $y_{-i} \in O_{\delta''}(a_{-i})$,
\begin{align*}
        &  u_i(t, \bm{b}_i^*(t_i), y_{-i})  \\
={}     &  \sum_{j=1}^J  \bigg[ \frac{\xi_i(\bm{b}_i^*(t_i), y_{-i})}{\xi_i(\bm{b}_i^*(t_i), y_{-i}) + \sum\limits_{\ell \ne i: y_\ell = \bar{a}} \xi_\ell(\bm{b}_i^*(t_i), y_{-i})} \psi_1(t^j_0, t_{11}, \ldots, t_{n1}, \bm{b}_i^*(t_i), y_{-i})  \\
        &  + \Big( 1 - \frac{\xi_i(\bm{b}_i^*(t_i), y_{-i})}{\xi_i(\bm{b}_i^*(t_i), y_{-i}) + \sum\limits_{\ell \ne i: y_\ell = \bar{a}} \xi_\ell(\bm{b}_i^*(t_i), y_{-i})} \big) \psi_2(t^j_0, t_{11}, \ldots, t_{n1}, \bm{b}_i^*(t_i), y_{-i}) \\
        &  + \varphi_i(t^j_0, t_{11}, \ldots, t_{n1}, \bm{b}_i^*(t_i), y_{-i}) \bigg]  \kappa(\{t^j_0\} \mid t) \\
\ge{}   &  \sum_{j=1}^J  \bigg[ \Big( \frac{\xi_i(\bm{b}_i(t_i), a_{-i})}{\xi_i(\bm{b}_i(t_i), a_{-i}) + \sum\limits_{\ell \ne i: a_\ell = \bar{a}} \xi_\ell(\bm{b}_i(t_i), a_{-i})} - \epsilon_1 \Big)  \psi_1(t^j_0, t_{11}, \ldots, t_{n1}, \bm{b}_i^*(t_i), y_{-i}) \\
        &  + \Big( 1 - \frac{\xi_i(\bm{b}_i(t_i), a_{-i})}{\xi_i(\bm{b}_i(t_i), a_{-i}) + \sum\limits_{\ell \ne i: a_\ell = \bar{a}} \xi_\ell(\bm{b}_i(t_i), a_{-i})} + \epsilon_1 \Big) \psi_2(t^j_0, t_{11}, \ldots, t_{n1}, \bm{b}_i^*(t_i), y_{-i}) \\
        &  + \varphi_i(t^j_0, t_{11}, \ldots, t_{n1}, \bm{b}_i^*(t_i), y_{-i}) \bigg]  \kappa(\{t^j_0\} \mid t) \\
>{}     &  \sum_{j=1}^J  \bigg[ (\frac{\xi_i(\bm{b}_i(t_i), a_{-i})}{\xi_i(\bm{b}_i(t_i), a_{-i}) + \sum\limits_{\ell \ne i: a_\ell = \bar{a}} \xi_\ell(\bm{b}_i(t_i), a_{-i})} - \epsilon_1) (\psi_1(t^j_0, t_{11}, \ldots, t_{n1}, \bm{b}_i(t_i), a_{-i}) - \tfrac{\epsilon}{3}) \\
        &  + \Big( 1 - \frac{\xi_i(\bm{b}_i(t_i), a_{-i})}{\xi_i(\bm{b}_i(t_i), a_{-i}) + \sum\limits_{\ell \ne i: a_\ell = \bar{a}} \xi_\ell(\bm{b}_i(t_i), a_{-i})} + \epsilon_1 \Big) (\psi_2(t^j_0, t_{11}, \ldots, t_{n1}, \bm{b}_i(t_i), a_{-i}) - \tfrac{\epsilon}{3}) \\
        &  + \varphi_i(t^j_0, t_{11}, \ldots, t_{n1}, \bm{b}_i(t_i), a_{-i}) - \tfrac{\epsilon}{3} \bigg]  \kappa(\{t^j_0\} \mid t) \\
={}     &  u_i(t, \bm{b}_i(t_i), a_{-i}) - \tfrac{2\epsilon}{3} \\
        &  - \epsilon_1 \sum_{j=1}^J  \Big[ \psi_1(t^j_0, t_{11}, \ldots, t_{n1}, \bm{b}_i(t_i), a_{-i}) - \psi_2(t^j_0, t_{11}, \ldots, t_{n1}, \bm{b}_i(t_i), a_{-i}) \Big]  \kappa(\{t^j_0\} \mid t) \\
>{}     &  u_i(t, \bm{b}_i(t_i), a_{-i}) - \epsilon.
\end{align*}
The first inequality holds since
\begin{align*}
\frac{\xi_i(\bm{b}_i^*(t_i), y_{-i})}{\xi_i(\bm{b}_i^*(t_i), y_{-i}) + \sum\limits_{\ell \ne i: y_\ell = \bar{a}} \xi_\ell(\bm{b}_i^*(t_i), y_{-i})}    &  \ge  \frac{\xi_i(\bm{b}_i^*(t_i), y_{-i})}{\xi_i(\bm{b}_i^*(t_i), y_{-i}) + \sum\limits_{\ell \ne i: a_\ell = \bar{a}} \xi_\ell(\bm{b}_i^*(t_i), y_{-i})}   \\
&  >  \frac{\xi_i(\bm{b}_i(t_i), a_{-i})}{\xi_i(\bm{b}_i(t_i), a_{-i}) + \sum\limits_{\ell \ne i: a_\ell = \bar{a}} \xi_\ell(\bm{b}_i(t_i), a_{-i})} - \epsilon_1,
\end{align*}
and
$$ \psi_1(t^j_0, t_{11}, \ldots, t_{n1}, \bm{b}_i^*(t_i), y_{-i})  \ge  \psi_2(t^j_0, t_{11}, \ldots, t_{n1}, \bm{b}_i^*(t_i), y_{-i}). $$
The other equalities and inequalities are due to simple algebras.

To summarize, in all the cases above, for any bidder~$i$ and all $(t, a_{-i})$, there exists a neighborhood $O_{a_{-i}}$ of $a_{-i}$ such that for all $y_{-i} \in O_{a_{-i}}$,
$$ u_i(t, \bm{b}_i^*(t_i), y_{-i}) - u_i(t, \bm{b}_i(t_i), a_{-i}) > - \epsilon. $$
That is, the all-pay auction game satisfies the condition of uniform payoff security.

\bigskip

Note that
\begin{align*}
\sum_{i \in I} u_i(t, a)  ={}   &  \sum_{i \in I}  \sum_{j=1}^J  v_i(t^j_0, t_1, \ldots, t_n, a_1, \ldots, a_n)  \kappa(\{t^j_0\} \mid t)    \\
                          ={}   &  \sum_{j=1}^J  \bigg[ \psi_1(t_0, t_{11}, \ldots, t_{n1},a) \\
                                & + (n - 1) \psi_2(t_0, t_{11}, \ldots, t_{n1}, a)  +  \sum_{i \in I} \varphi_i(t_0, t_{11}, \ldots, t_{n1}, a) \bigg] \kappa(\{t^j_0\} \mid t),
\end{align*}
which is continuous in $a$ for every $t \in T$. By Lemma~\ref{lem:secure}, the game possesses a behavioral-strategy equilibrium.

\bigskip

Recall that $\cT_i = \cB(T_i)$, the Borel $\sigma$-algebra on $T_i$. Let $\cF_i = T_i^{\cB(\bR) \otimes \{\emptyset, \bR\}}$; that is, $\cF_i$ is the restriction of the $\sigma$-algebra $\cB(\bR) \otimes \{\emptyset, \bR\}$ to $T_i$, where the trivial $\sigma$-algebra $\{\emptyset, \bR\}$ is imposed on the dimension $t_{i2}$ for each $i \in I$. Then $\cT_i$ is nowhere equivalent to $\cF_i$. Due to Corollary~\ref{coro-interdependence}, DCPI is satisfied. By Proposition~\ref{prop:discontinuous}, a pure-strategy equilibrium exists.
\end{proof}

\singlespacing

\end{document}